\documentclass[a4paper,noindent]{JHEP}
\usepackage{amsmath,graphicx,amssymb,multirow,bbm}

\usepackage{epsfig, subfigure}
\usepackage{slashed}
\usepackage[english]{babel}

\def\ga{\mathrel{\raise.3ex\hbox{$>$\kern-.75em\lower1ex\hbox{$\sim$}}}}
\def\la{\mathrel{\raise.3ex\hbox{$<$\kern-.75em\lower1ex\hbox{$\sim$}}}}

\newcommand{\lam}{\lambda}
\def\lsim{\mathrel{\rlap{\lower4pt\hbox{\hskip1pt$\sim$}}
    \raise1pt\hbox{$<$}}}                
\def\gsim{\mathrel{\rlap{\lower4pt\hbox{\hskip1pt$\sim$}}
    \raise1pt\hbox{$>$}}}                

\usepackage{cite}

\makeatletter
\renewcommand\appendix{\par
  \setcounter{section}{0}%
  \setcounter{subsection}{0}%
  \renewcommand\thesection{\@Alph\c@section}
  \setcounter{figure}{0}%
  \setcounter{table}{0}%
  \renewcommand\thefigure{\thesection.\@arabic\c@figure}
  \renewcommand\thetable{\thesection.\@arabic\c@table}
  }
\makeatother

\title{Charged Higgs bosons in single top production at the LHC}

\author{
Renato Guedes\\
Centro de F\'\i sica Te\' orica e Computacional, Faculdade de Ci\^encias, Universidade de Lisboa, Av. Prof. Gama Pinto 2, 1649-003 Lisboa, Portugal.\\
Email: \email{renato@cii.fc.ul.pt}}
\author{
Stefano Moretti\\
School of Physics and Astronomy, University of Southampton, Highfield, Southampton SO17 1BJ, UK.\\
Email: \email{stefano@phys.soton.ac.uk}}
\author{
Rui Santos\\
Instituto Superior de Engenharia de Lisboa, Rua Conselheiro Em\'\i dio Navarro 1, 1959-007 Lisboa, Portugal \& \\
Centro de F\'\i sica Te\' orica e Computacional, Faculdade de Ci\^encias, Universidade de Lisboa, Av. Prof. Gama Pinto 2, 1649-003 Lisboa, Portugal.\\
Email: \email{rsantos@cii.fc.ul.pt}}

\abstract{We show that a light charged Higgs boson signal via $\tau^\pm\nu$ decay 
can be established at the Large Hadron Collider (LHC) 
also in the case of single top production. This process complements 
searches for the same 
signal in the case of charged Higgs bosons emerging from  $t \bar t$ 
production. The models accessible include the Minimal Supersymmetric Standard
Model (MSSM) as well a variety of 2-Higgs Doublet Models (2HDMs). High energies and 
luminosities are however required, thereby restricting interest on this mode to the
case of the LHC running at 14 TeV with design configuration.}  

\preprint{SHEP-12-02}

\begin{document}


\section{Introduction}

\indent
The Large Hadron Collider (LHC) is presently operating at a Centre-of-Mass (CM) energy of 8 TeV, following the 
highly successful stage at 7 TeV, where about 5 fb$^{-1}$ of luminosity 
were collected. The expectation at
present for the current run is a 15 fb$^{-1}$ data sample. In a few years from now, the LHC will be operating
at design energy and luminosity, i.e., at 14 TeV and with the prospect of gathering of the order of 300
fb$^{-1}$ or so of data. The highest priority of the CERN machine is to finally confirm, or indeed disprove,
the current paradigm for Electro-Weak Symmetry Breaking (EWSB), whereby a Higgs mechanism is postulated to give
mass to force and matter states by spontaneously breaking the underlying SU(2)$_L\otimes$U(1)$_Y$ gauge symmetry
(involving isospin $L$ and hypercharge $Y$) while instead preserving SU(3)$_C$ (where $C$ is colour).
While it was first embedded within the Standard Model (SM), wherein the Higgs mechanism is implemented
in its minimal form, i.e., through a single Higgs doublet complex field (from which then one scalar
Higgs field finally emerges after EWSB), it is now clear that Beyond the SM (BSM) physics is required,
given overwhelming experimental evidence that cannot be explained within the SM alone (like neutrino masses,
Dark Matter (DM), the matter-antimatter asymmetry, etc.). Conversely, recent LHC
evidence of possible neutral Higgs signals at a mass of 124--126 GeV \cite{ATLAS:2012ae,Chatrchyan:2012tx},
compatible with the SM hypothesis, with a slight excess in the $h \to \gamma \gamma$ channel, 
calls for not dismissing the Higgs mechanism as the
means for achieving EWSB. So that one may well argue that exploring the possibility of BSM physics,
still reliant on the Higgs hypothesis for EWSB yet with the latter realised in some non-minimal version, has become
the main focus at present of the phenomenological activities concerned with LHC data.

\indent
One of the most striking evidences of BSM physics employing a non-minimal Higgs mechanism as the source of EWSB
would be the appearance of a (singly) charged Higgs boson, $H^\pm$. The latter is predominantly (but not only) produced
in top-(anti)quark decays, so long that it is light, i.e., $m_{H^\pm}<m_t$ (the top mass). Presently,
the LEP experiments have set a lower limit on the mass of a charged Higgs boson, of 79.3 GeV at 95\% Confidence Level (CL), 
assuming that BR$(H^+ \to \tau^+ \nu) + BR(H^+ \to c \bar s)=1$ holds for the possible charged Higgs boson
Branching Ratios (BRs)~\cite{LEP}. This limit becomes stronger 
if BR$(H^+ \to \tau^+ \nu)  \approx 1$ (see \cite{Logan:2009uf} for a discussion). Searches at 
the Tevatron~\cite{Tev} based on $t \bar t$ production with one of the tops decaying via $t \to b H^+$ and assuming 
BR$(H^+ \to \tau^+ \nu) \approx 1$ have yielded a limit of BR$(t \to b H^+) < 0.2$ for a charged Higgs mass 
of 100 GeV.  The LHC is currently providing limits similar to those obtained at the
Fermilab machine \cite{Hpm-LHC}.

\indent
The simplest extensions of the SM that give rise to charged Higgs bosons amount to the addition of an 
extra Higgs doublet to the SM field content. The most common CP-conserving 2HDM  has a softly broken 
$Z_2$ symmetry. When this symmetry is extended to the fermions to avoid Flavour Changing Neutral Currents (FCNCs) 
we end up with four~\cite{barger} different models, which we will call Type I, 
Type II, Type Y and Type X~\cite{KY}  (named I, II, III and IV in~\cite{barger}, respectively). Constraints 
from $B$-physics, and particularly those coming from $b \to s \gamma$~\cite{bsgamma}, have excluded a charged 
Higgs boson with a mass below approximately 300 GeV almost independently of $\tan \beta = v_2/v_1$ -- the ratio of the 
Vacuum Expectation Values (VEVs) of the two doublets -- in models Type II and Type Y\footnote{However,
if a 2HDM Type II is embedded in Supersymmetry, such a realisation of EWSB remains possible, as the
additional sparticle states available, e.g., in the MSSM, can cancel out the $H^\pm$ contributions
to $B$-physics observables.}. Charged Higgs bosons with masses as low as 
100 GeV are instead still allowed in models Type I and Type X~\cite{Logan:2009uf,KY,Su:2009fz}. 

\indent
These scenarios as well as their experimental and theoretical  constraints have been discussed in detail in 
\cite{Aoki:2011wd}, to which we refer the reader also for conventions and notation. In that paper, which 
reviewed the LHC scope in probing 2HDMs through the detection of a light $H^\pm$ state in all possible
production modes, predominantly decaying via $H^\pm\to\tau^\pm\nu$, one channel was singled out as offering 
clear prospects of detection, i.e., $H^\pm$ production from top-(anti)quark decays where the latter is produced
in single mode, as opposed to the case wherein the top-(anti)quark is produced in pairs, as it was customary to exploit 
(in fact, successfully) in previous literature. This conclusion was strongly supported in that paper by a detailed 
parton level analysis, however, the authors of \cite{Aoki:2011wd} also cautioned that this result should have eventually been put
on firmer ground through a full Parton Shower (PS), hadronisation and detector study. Only after this, the yield 
of the `single-top' channel could be contrasted with the one obtained through the `double-top' mode and the exclusion and discovery reaches
of either channel relatively assessed. 

\indent
It is the aim of this paper to carry out this PS, hadronisation and detector analysis and to attempt comparing
the ensuing new results from `single-top' with the time honoured ones established through `double-top' analyses. 
In this study we will concentrate on  the 14 TeV CM energy because, as it
will become clear later on, not only the cross section for the single-top mode
grows significantly with energy (in view of its dominant $t$-channel topology)
but also a large amount of luminosity is needed to start probing it at statistically profitable levels. 
The analysis will be carried out in the context of 2HDMs, however, it is
important to note that the phenomenology of a light charged Higgs boson as discussed herein
is much more general. There is in fact a large number of BSM Higgs scenarios
that share a common charged Higgs boson phenomenology for vast regions of their parameter space with those
discussed here.  All such models have in common the fact they have a specific type of 2HDM as constituent for 
achieving EWSB. Recently, a number of these scenarios have been discussed in the literature~\cite{Aoki:2008av}, wherein the 
charged Higgs boson BR into leptons (including $\tau$'s) is enhanced relative to the SM case (Type X),
which provide DM candidates naturally and can finally accommodate neutrino oscillations as well as the strong first order phase transition 
required for successful baryogenesis while being in agreement with all experimental data.

\indent
The plan of the paper is as follows. The next section 
describes the main production and decay modes of such a light charged Higgs state at the LHC alongside the corresponding backgrounds. The following one 
documents our selection strategy and presents our main results. 
Finally, we conclude in Section 4.

\section{$H^\pm$ Signal and SM backgrounds}
\label{sec:signal}

Although the single-top channel is not the main top-(anti)quark production process at the LHC, 
it is still significant enough to deserve a full investigation regarding its
contribution to the production of charged Higgs bosons. We
will mainly focus on a light charged Higgs boson produced via $t$-channel graphs,
$pp \to t  \,  j \to H^+ \, \bar  b \, j $ and $H^+  \to   \tau^+ \, \nu $,
together with 
$pp \to \bar t  \,  j \to H^- \,  b \, j $ and $H^-  \to   \tau^- \, \bar \nu $,
where $j$ represents a light-quark jet.
In what follows we are considering proton-proton collisions
at a center-of-mass (CM) energy of  $\sqrt{s} = 14$ TeV and a top-quark
mass $m_t = 173$ GeV. The theoretical normalisations presented in this
section for the different single-top production processes are all 
through Next-to-Next-to-Leading-Order (NNLO) 
accuracy, albeit limited to the Next-to-Next-to-Leading-Logarithmic (NNLL) component, as described 
in~\cite{Kidonakis:2010ux,Kidonakis:2010tc,Kidonakis:2011wy}. We have used higher order normalisations in presenting
the forthcoming signal results, as described below.

\indent
There are three distinct contributions to single top-production
at the LHC. We first consider $tW$ associated production which, at lowest order, is the
sum of the two partonic processes $bg \to t \, W^-$ and 
$\bar bg \to \bar t \, W^+$. The cross section for $\sqrt{s} = 14$ TeV
at NNLO and for $m_t = 173$ GeV is given by~\cite{Kidonakis:2010ux}
\begin{equation}
\sigma^{\rm NNLO}_{\bar t W^+}  ={41.8 \, \pm \, 1.0 \, }^{+1.5}_{-2.4} \quad {\rm pb}
\label{eq:tW}
\end{equation}
where the first uncertainty is from the renormalisation/factorisation
scale variation between $m_t/2$ and $2m_t$ 
and the second is from the MSTW2008 NNLO 
Parton Distribution Functions (PDFs)~\cite{Martin:2009iq} at 90\% CL.
The cross section is the same for both top and antitop production and therefore
the sum of the two processes yields 83.6 pb.
Second, we consider $s$-channel production via $q \, \bar q' \to  t \, \bar b$
and $\bar q \, q' \to  \bar t\, b$. The cross sections were calculated in~\cite{Kidonakis:2010tc}:
\begin{equation}
\sigma^{\rm NNLO}_{t \bar b}  ={7.93 \, \pm \, 0.14 \, }^{+0.31}_{-0.28} \quad {\rm pb}
\end{equation}
and 
\begin{equation}
\sigma^{\rm NNLO}_{\bar t b}  ={3.99 \, \pm \, 0.05  \, }^{+0.14}_{-0.21} \quad {\rm pb}
\end{equation}
where uncertainties are as described for eq.~(\ref{eq:tW}). The total cross section
for top plus antitop production via $s$-channel is then 11.9 pb.
However, the most important contribution for single-top at the LHC
comes from the aforementioned $t$-channel process. The total cross section for 
$t$-channel top production is in fact \cite{Kidonakis:2011wy}
\begin{equation}
\sigma^{\rm NNLO}_{t \, j}  ={151 \,}^{+4}_{-1} \, \pm \, 4  \quad {\rm pb}
\end{equation}
while for antitop we have
\begin{equation}
\sigma^{\rm NNLO}_{\bar t j}  ={92 \,  }^{+2 \, +2}_{-1 \, -3} \quad {\rm pb}
\end{equation}
where again the first uncertainty is from scale variation
and the second is from the PDFs. The total
cross section in $t$-channel is therefore 243 pb\footnote{By comparison, note that 
the NLO cross section for $t\bar t$ production at the LHC is about 800 pb.}.

\indent
The signal consists of the sum of all possible single-top production processes
with the subsequent decays $t \to H^+ \, b  \to   \tau^+ \, \nu \,  b$
plus all the antitop counterparts with decays 
$\bar t \to H^- \, \bar b  \to   \tau^- \, \bar{\nu} \, \bar b$.
Taking into account the LEP bounds on the mass of a charged Higgs
boson, we consider a mass ranging from 90 to 160 GeV and the analysis
is performed in 10 GeV mass steps. In order to maximise the signal-to-background significance
($S/\sqrt B$), 
it turns out that both the $s$-channel and the $tW$ single-top production
modes  become negligible. Therefore, when discussing 
the signal, it is in fact to the $t$-channel process that we will be referring to,
as it is the only one that survives the set of cuts imposed.
The signal (single-top) events were generated with POWHEG~\cite{Alioli:2010xd}
at NLO with the CTEQ6.6M~\cite{Nadolsky:2008zw} PDFs.
The top was then decayed in PYTHIA~\cite{Sjostrand:2006za}. We have considered
only the leptonic decays of the tau-leptons, that is, the signal final state is
$pp \to l \, b \, j \, \slashed{E}$, where $l=e, \mu$ (electrons and muons)
while $\slashed{E}$ means missing (transverse) energy.

\indent
The irreducible background to this process is obviously single-top production
with the subsequent decay $t \to b \, W^+$. Again, all single-top
production processes, also generated with POWHEG, were included.
We have taken into account the most relevant contributions
to the reducible background too: i.e., $t\bar t$ production, $W^\pm ~+$~jets (including not only light quarks and gluons,
but also $c$- and $b$-quarks) 
and the pure QCD background ($jjj$, where $j$ is any jet), which we will now
discuss in turn. The $t\bar t$ background was generated
with POWHEG including the top and $W^\pm$ boson decays
to all possible final states. The $W^\pm$ + jets (1, 2 and 3 jets) noise was
generated with  AlpGen~\cite{Mangano:2002ea} with CTEQ6ll PDFs.
Finally, the QCD background was generated with 
CalcHEP~\cite{Pukhov:2004ca} and, as before, the CTEQ6ll PDFs were used.

\indent
The hadronisation was performed with PYTHIA 6. The Perugia
tune~\cite{Skands:2009zm}
was used to handle the underlying events in POWHEG
while the ATLAS MC09 tune~\cite{ATLAStune} was used for events
generated with AlpGen and CalcHEP. We also have cross-checked
the single-top event generation with AcerMC~\cite{Kersevan:2004yg}  and 
the $W^\pm$ + jets events were also generated with MCFM~\cite{Campbell:1999ah, site}
(whose normalisation we have used).
To avoid double counting in the $W^\pm$ + jets background, we have used the 
MLM matching scheme~\cite{mlm}. Following~\cite{mlm}, we 
have used the following cuts in generation: all jets have transverse 
momentum above 20 GeV, $|\eta_j|<2.5$ and $\Delta R_{jj}>0.7$ while
the missing (transverse) energy has to be larger than 20 GeV\footnote{We also have (slightly) changed the cuts at generation level to make sure
that no systematic bias was introduced by these, by confirming that only the efficiency of generating the ensuing hadronic final states 
changed, not
their yields.}.

\indent
After hadronisation, DELPHES~\cite{Ovyn:2009tx}, 
which is a framework for the fast simulation of a 
generic collider experiment, was used 
to simulate the detector effects.
For the detector and trigger configurations,  we resorted to the ATLAS 
default definitions.

\section{Selection}

In this section we describe the analysis and the selection cuts in detail. First we have used the trigger card in ATLAS
which for our purposes means that we have asked for an isolated electron with $p_T > 25$ GeV
or an isolated muon with $p_T > 20$ GeV. We will now describe the analysis in detail and we start
by noting that the pure QCD background is not shown in the plots due to its (initially) overwhelming magnitude. However,
we will highlight the selection cuts that have eliminated this background. 

\begin{figure}[h!]
  \begin{center}
    \includegraphics[scale=0.37]{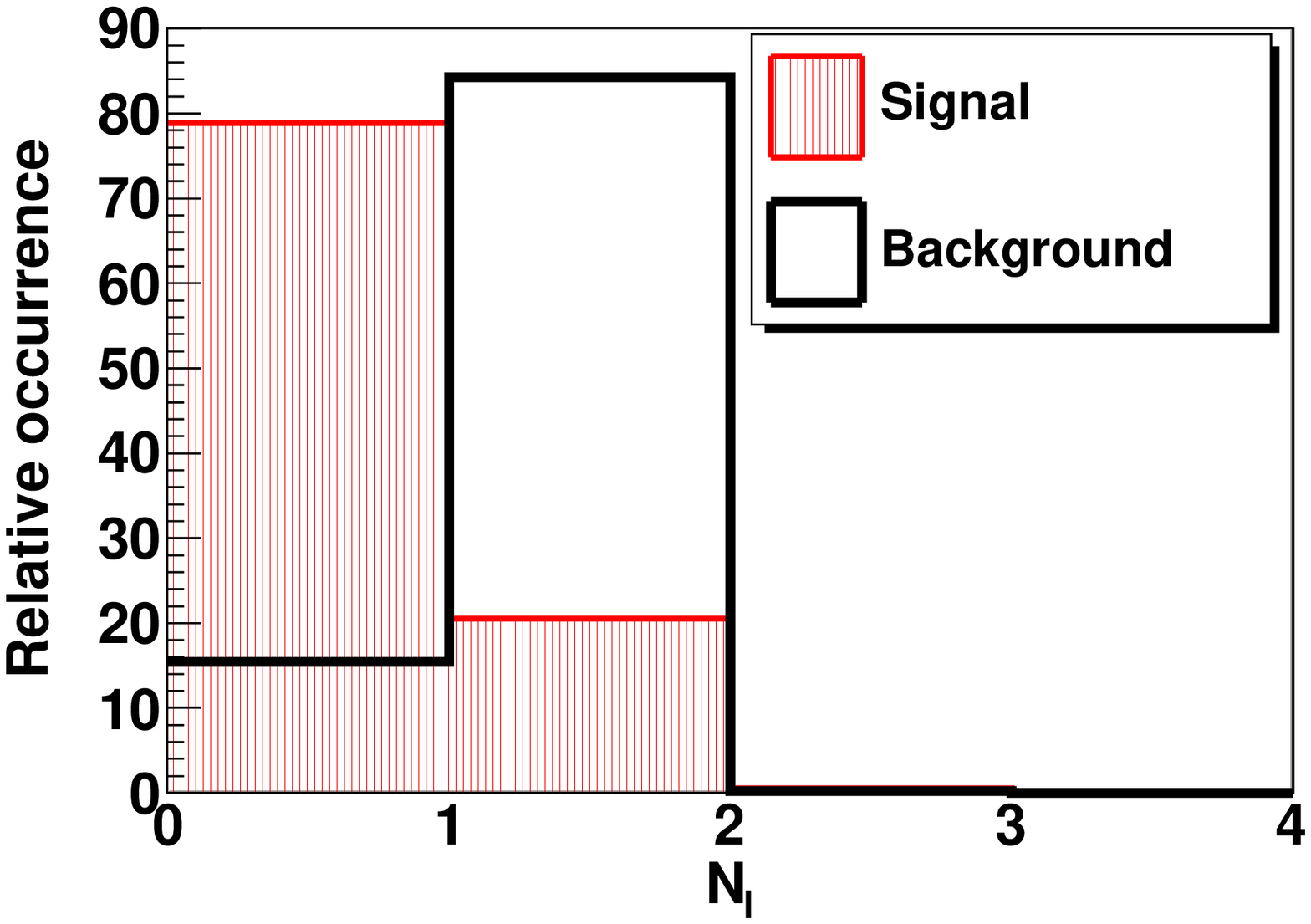}
    \includegraphics[scale=0.37]{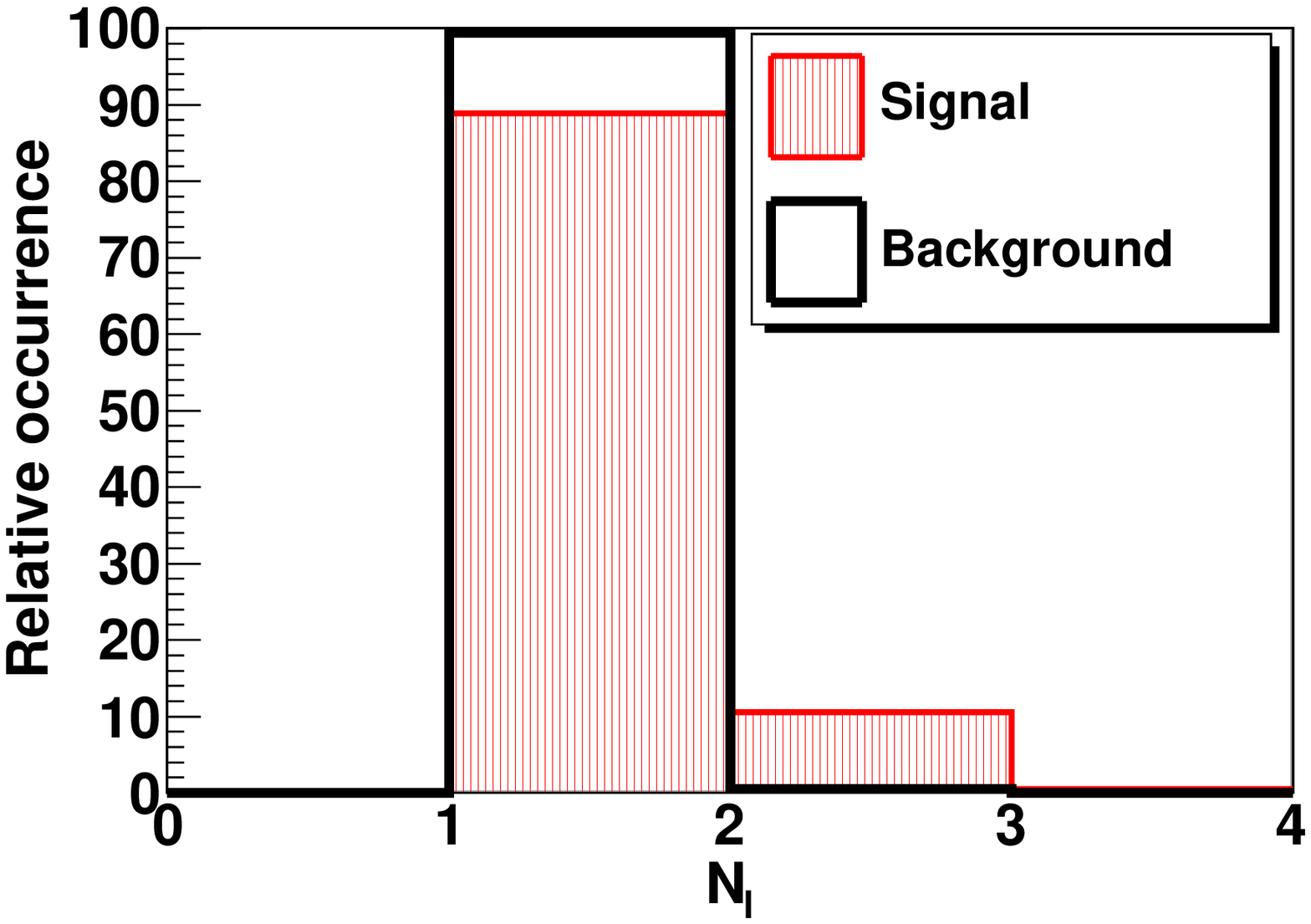}
    \caption{In the left panel we present the number of muons with $p_T > 20$ GeV or the number of electrons 
    with $p_T > 30$ GeV before the first cut. On the right we show the number of leptons with a 
    transverse momentum of at least 10 GeV passing the first cut.}
    \label{fig1}
  \end{center}
\end{figure}

\begin{enumerate}

\item We demand one electron with transverse momentum larger  than 30 GeV or a muon with transverse momentum above 20 GeV. 
In both cases, the lepton has to be in the central region of the detector, $|\eta| < 2.5$. In the left panel of figure~\ref{fig1} 
we show the leptonic multiplicity before these cuts.

\item We veto events with two or more leptons with transverse moment above 10 GeV.
In the right panel of figure~\ref{fig1} we show the leptonic multiplicity before such a veto. Although
the figure seems to indicate that events with two leptons in the final state should be retained,
the subsequent cuts show that the background would rise sharply above the signal, should this have been the case. This
cut eliminates the leptonic $t \bar t$ background almost completely.

\item In figure~\ref{fig2} (left) we show the lepton transverse momentum distribution after imposing
the first two cuts. Due to overwhelming W+0 jets background, it is not clear that
the best choice to maximise the sensitivity is to exclude events with leptons having $p_T$ above 55 GeV.
\begin{figure}[h!]
  \begin{center}
    \includegraphics[scale=0.37]{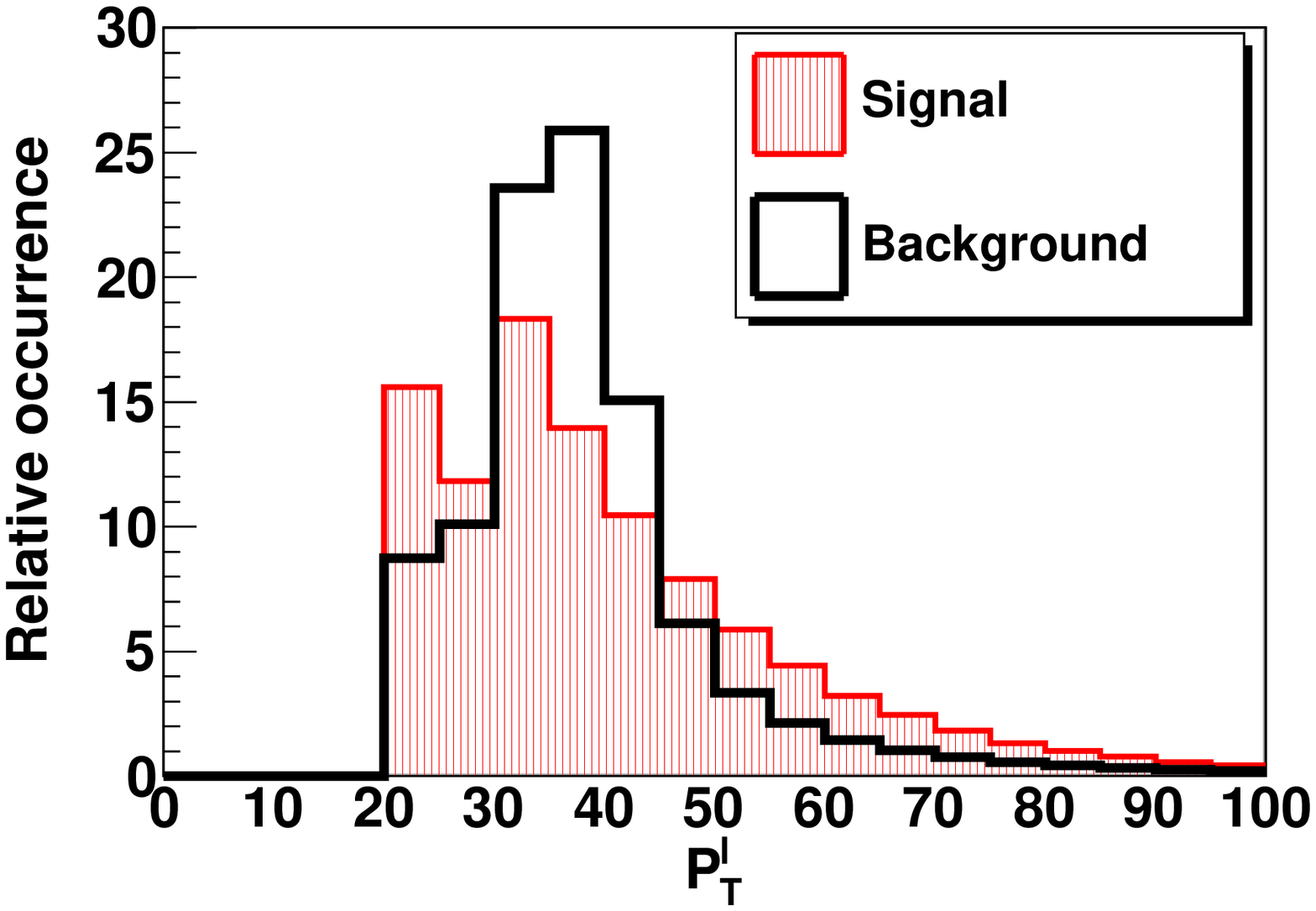}
    \includegraphics[scale=0.37]{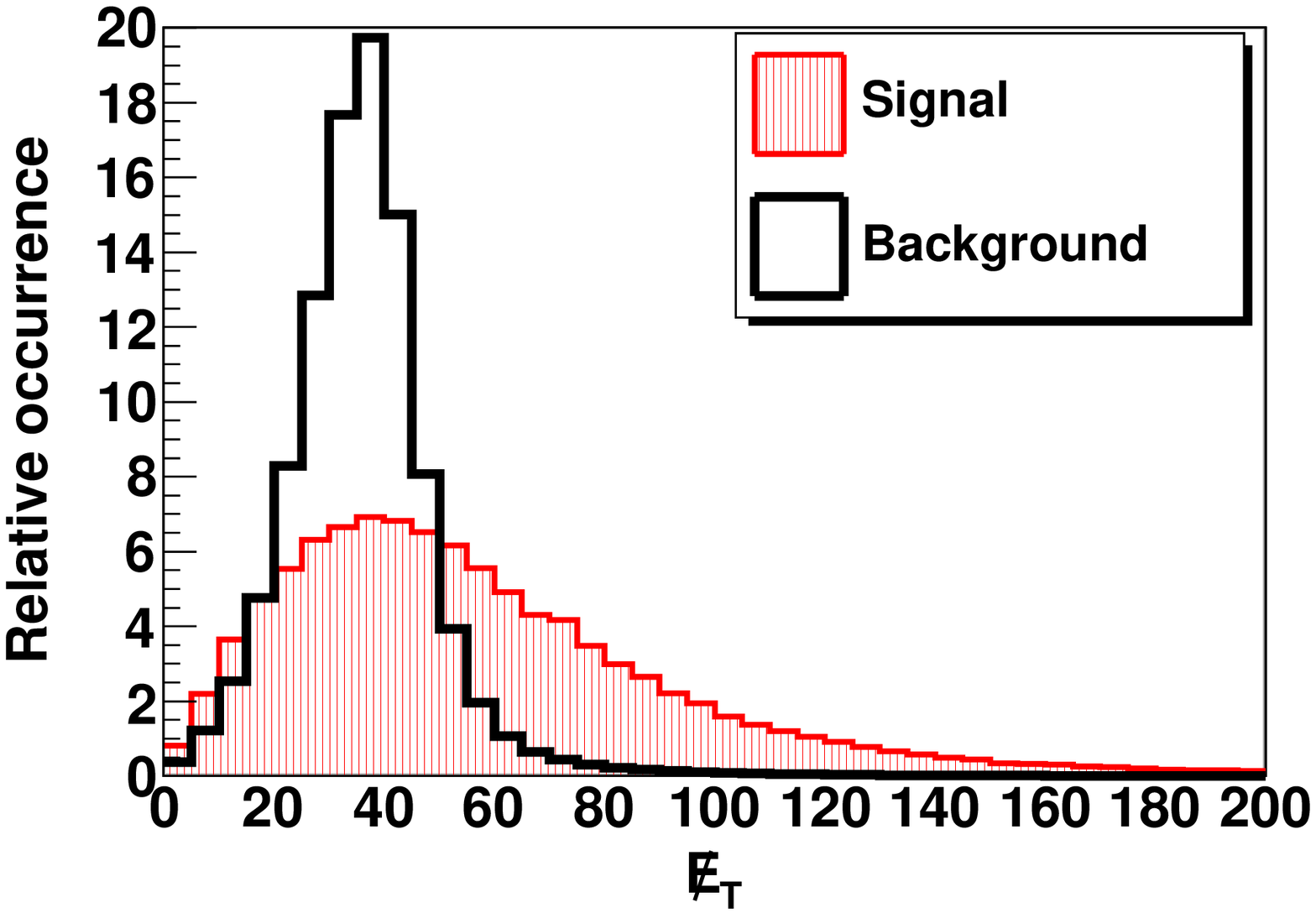}
    \caption{In the left panel the leptons transverse momentum distribution is shown, after imposing both the first and second cut.
    In the right panel we present the missing transverse energy distribution after the first two cuts. }
    \label{fig2}
  \end{center}
\end{figure}

\item  In the right panel of figure~\ref{fig2} we present the missing transverse energy distribution after the first two cuts.
We have therefore imposed that events with missing energy below 50 GeV should be excluded. This is another cut that 
dramatically reduces the QCD background.

\item In figure ~\ref{fig3} (left) we display the number of $b$-tagged jets after the first four cuts while in the right panel we show the $b$-tagged
jets transverse momentum distribution. We ask for one and only one $b$-tagged jet with a transverse momentum below 75 GeV. Again, although
two $b$-tagged jets have a higher relative occurrence in signal events, the latter drops significantly once the subsequent cuts
are imposed.  We assume for each $b$-tagging efficiency of a $b$-quark jet the value $R=0.7$, while for
the case of $c$-quark jets we take 0.1 and for light-quark/gluon jets we adopt 0.01. 
The background in question is the semi-leptonic $t \bar t$ background.

\begin{figure}[here]
  \begin{center}
    \includegraphics[scale=0.37]{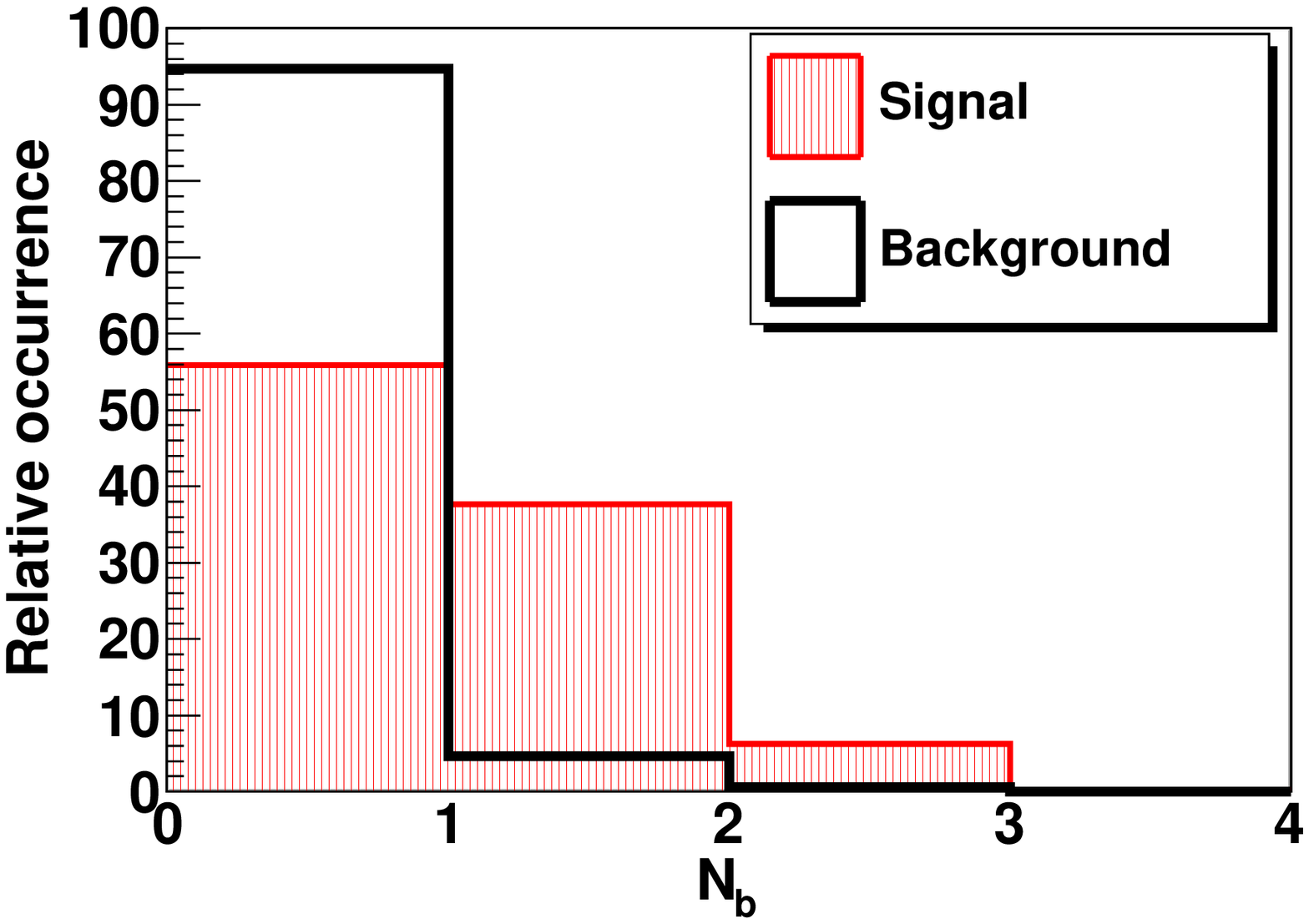}
    \includegraphics[scale=0.37]{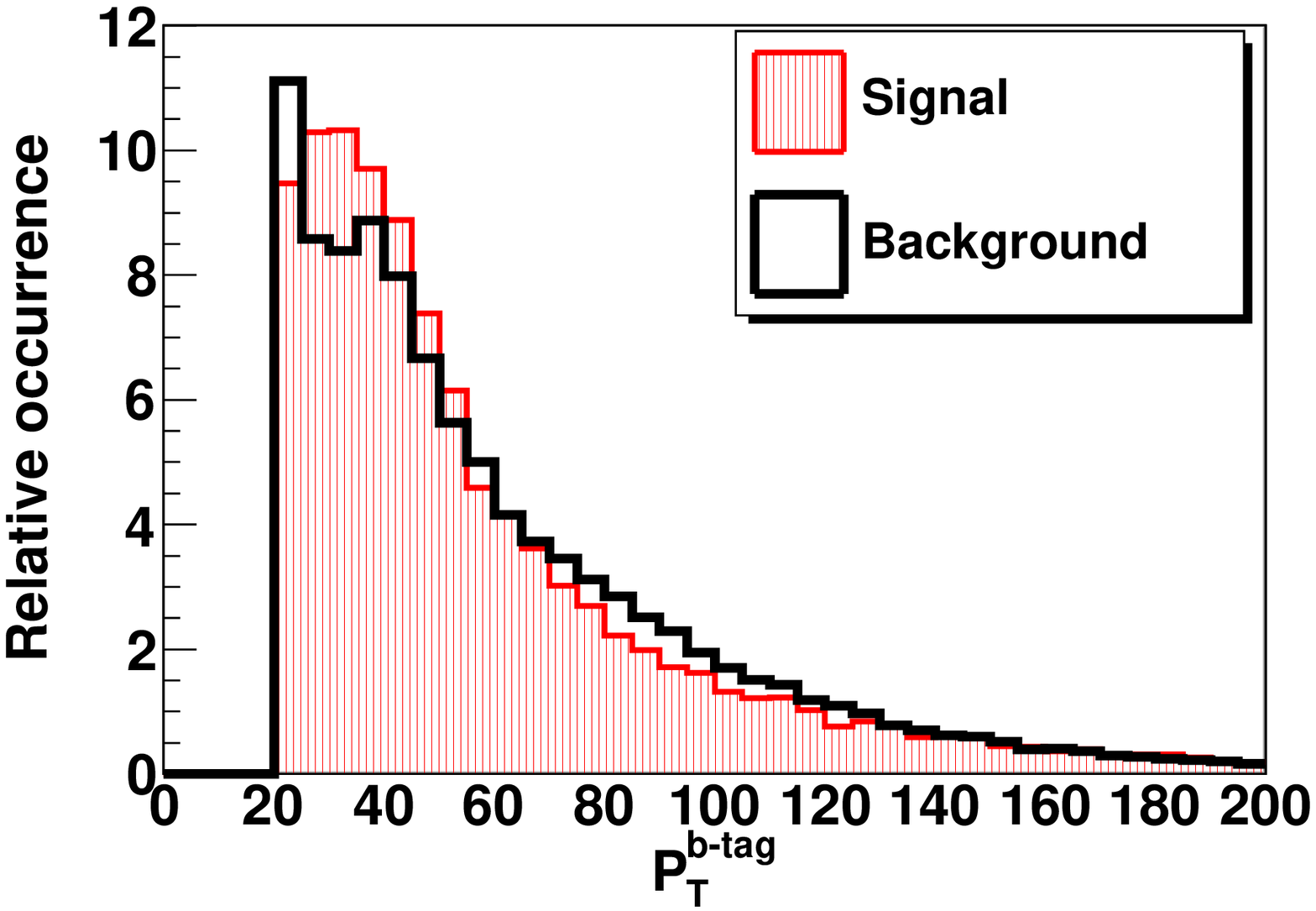}
    \caption{The $b$-jet selection cuts: number of  $b$-tagged jets with at least 30 GeV (left) and $b$-tagged jet transverse momentum distribution (right).}
    \label{fig3}
  \end{center}
\end{figure}

\begin{figure}[!htbp]
  \begin{center}
    \includegraphics[scale=0.37]{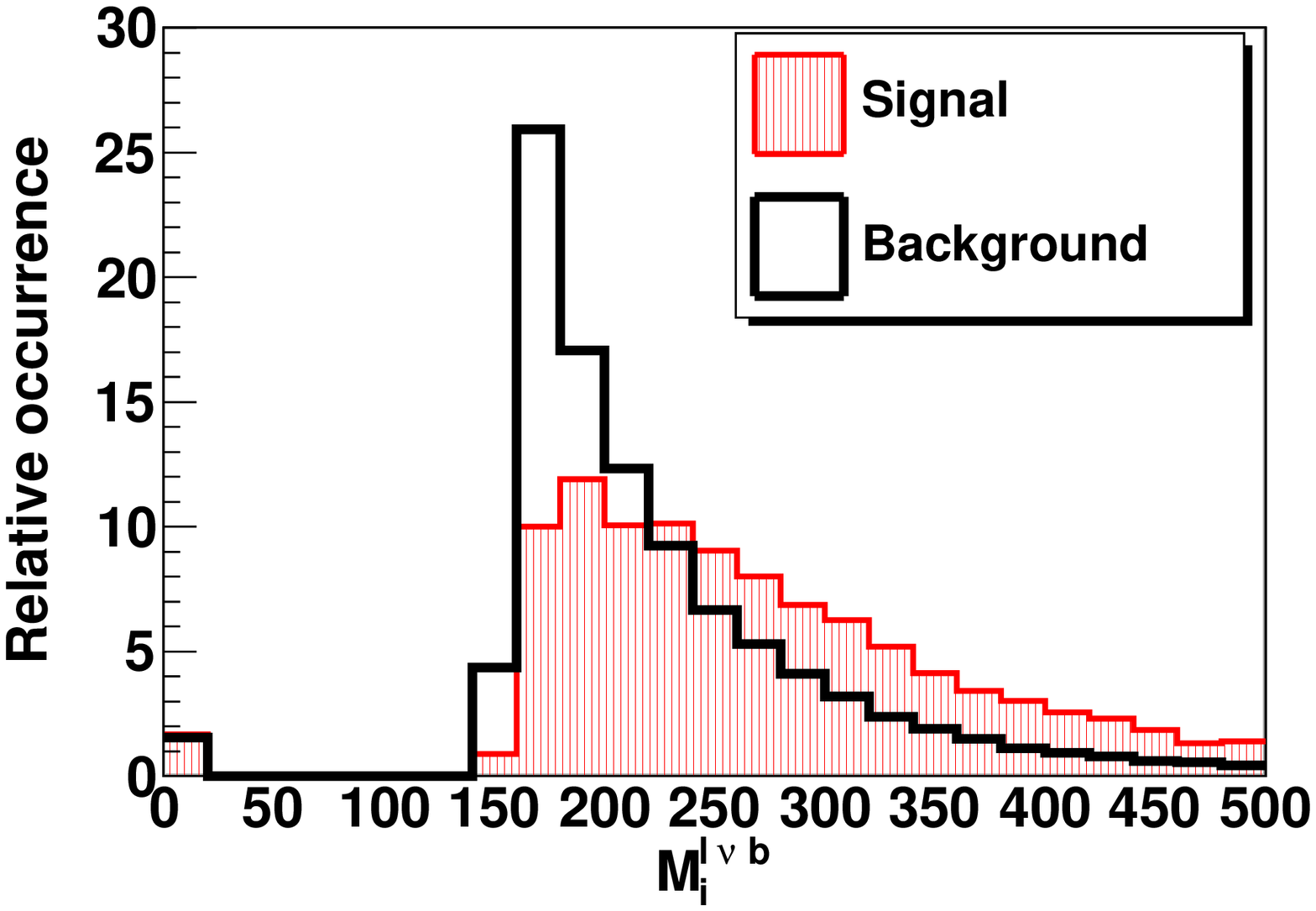}
    \includegraphics[scale=0.37]{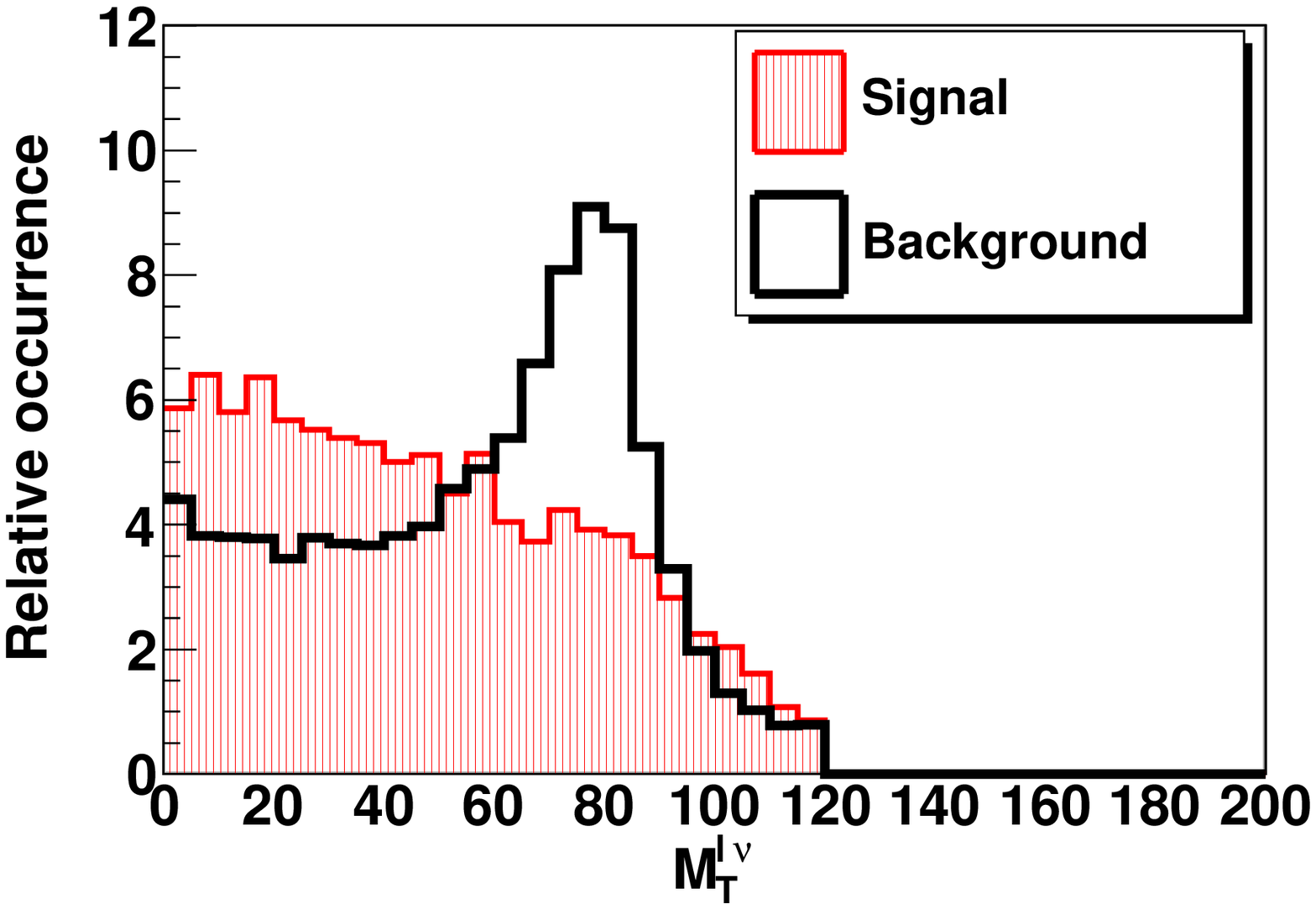}
    \caption{The ``top quark invariant mass'' distribution (left) and leptonic transverse invariant mass distribution (right).}
    \label{fig4}
  \end{center}
\end{figure}

\item In order to reconstruct the top mass we proceed in the following way. As we do not know the longitudinal component of the missing energy,
 we look for that component of the missing energy that together with the lepton momentum reconstructs the charged Higgs mass according to the relation
\begin{equation}
\slashed p_z=\frac{-b\pm\sqrt{b^2-4ac}}{2a},
\label{eq:pz}
\end{equation}
where
\begin{eqnarray}
a & = & \left(\frac{p_{zl}}{p_l}\right)^2-1 ,\\
b & = & 2\left(\frac{p_{xl}\slashed p_x + p_{yl}\slashed p_y}{p_l}+\frac{m_{H^+}^2}{2p_l}\right)\frac{p_{zl}}{p_l}, \\
c & = & \left(\frac{p_{xl}\slashed p_x + p_{yl}\slashed p_y}{p_l}+\frac{m_{H^+}^2}{2p_l}\right)^2 - \slashed p_T^2.
\end{eqnarray}
Here, the $\slashed p_i$'s are the missing energy momentum components, $p_{il}$ are the lepton momentum components, $p_{l}$ is the 
(massless) lepton energy and $\slashed p_T$
is the transverse missing energy.
With this information we can now reconstruct a ``top quark invariant mass''. Of course, because we will reconstruct fake top masses,
there is a wide distribution that includes unphysical masses. This distribution is presented in the left panel of figure~\ref{fig4}.
Whenever we obtained an imaginary value for $\slashed p_z$, the corresponding events were placed in the $0 - 20$ GeV bin. 
Between the two solutions of
eq.~(\ref{eq:pz}) we have chosen the one that gave a ``top quark invariant mass'' closest 
to the top quark mass experimental value. Finally, we 
demand events to have a ``top quark invariant mass'' above 280 GeV.


\item Next, we define the leptonic transverse mass~\cite{tinv} as
\begin{equation}
M_T^{l \nu}=\sqrt{2 p_{Tl} \slashed{p}_T-2 (p_{xl} \slashed{p}_x+p_{yl} \slashed{p}_y)},
\end{equation}
where $p_{Tl}$ is the lepton transverse momentum. The leptonic transverse mass distribution is
shown in figure~\ref{fig4} (right). To maximise the significance we have accepted events with 
$30 \, {\rm GeV} < M_T^{l \nu}  < 60 \, {\rm GeV} $ for charged Higgs masses between 90 and 130 GeV and 
$30 \, {\rm GeV} < M_T^{l \nu}  < 60 \, {\rm GeV} $ or $M_T^{l \nu} > 85 \, {\rm GeV}$ 
for higher values of the charged Higgs mass.
Again, figure~\ref{fig4} (right) would lead us to include values of $M_T^{l \nu} $ below 30 GeV.
However, we have excluded events with $M_T^{l \nu}  < 30 \, {\rm GeV} $ to further reduce the pure QCD background.

\begin{figure}[here]
  \begin{center}
    \includegraphics[scale=0.37]{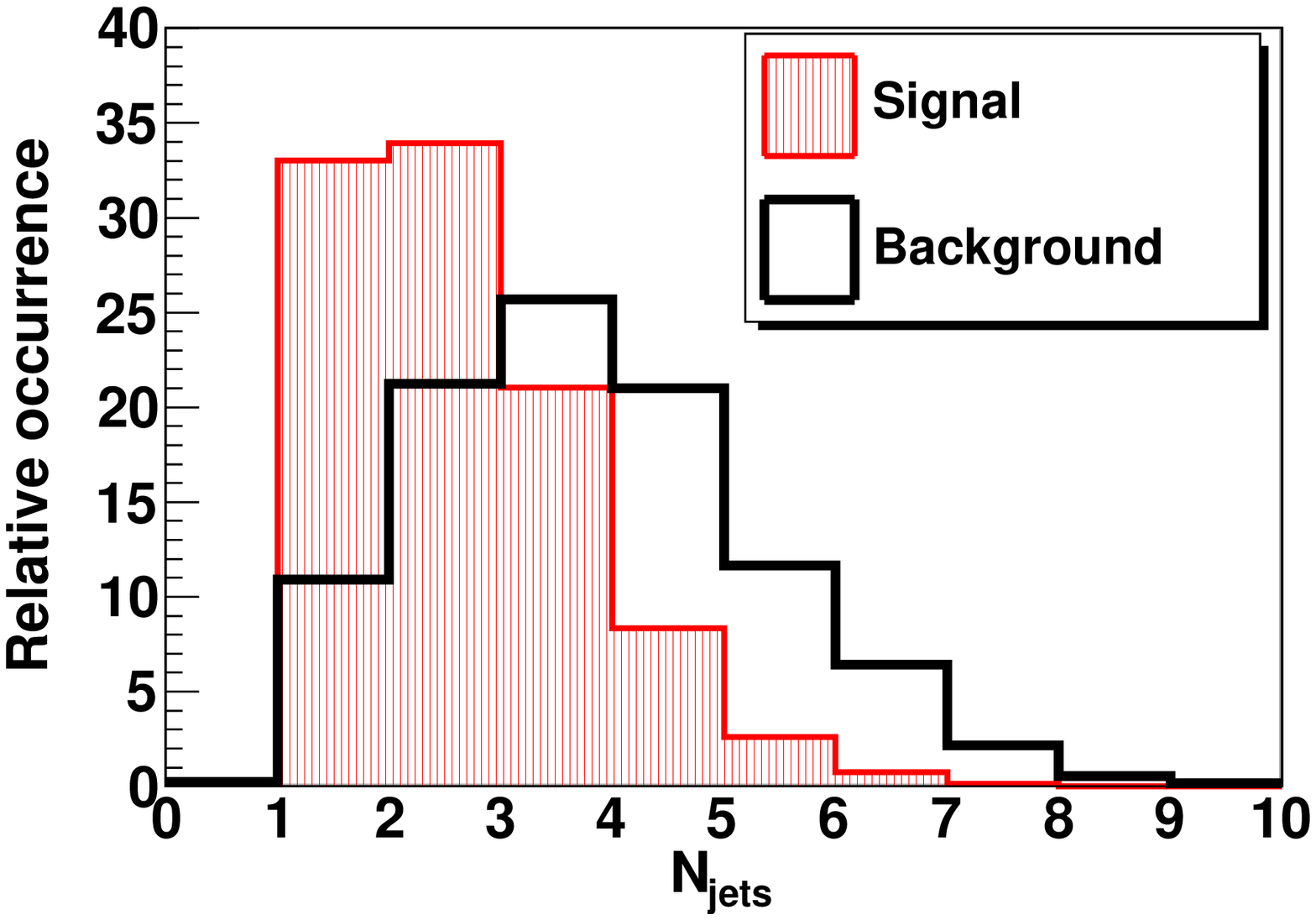}
    \includegraphics[scale=0.37]{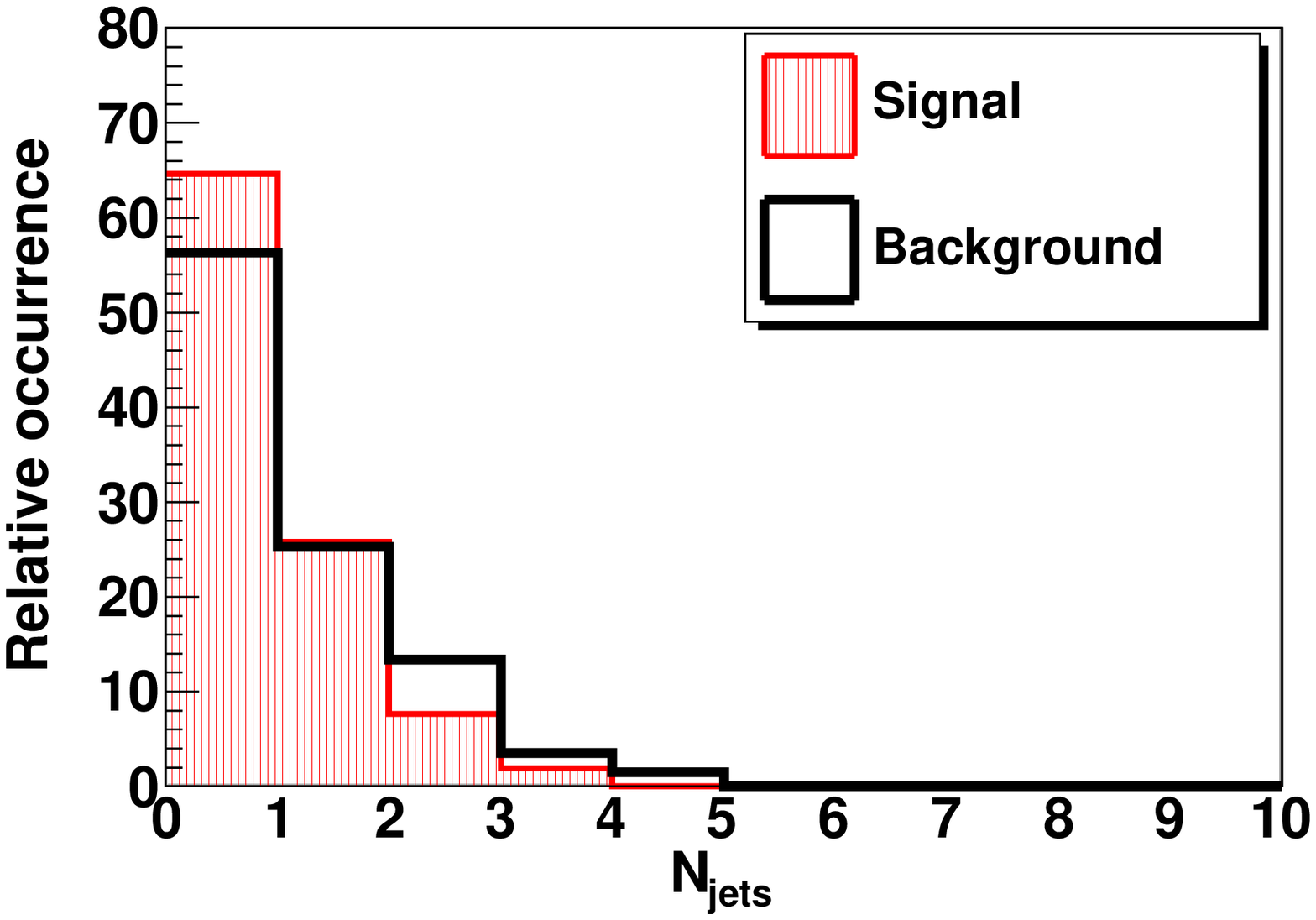}
    \caption{On the left we present the number of jets with transverse momentum above 30 GeV. 
    On the right we show the number of jets that passed the previous cut and have one more jet
     with a transverse momentum between 15 GeV and 30 GeV. In both cases we require the jet pseudorapidity
to be less the 4.9 in absolute value. }
    \label{fig5}
  \end{center}
\end{figure}

\item In the left panel of figure~\ref{fig5} we present the jet multiplicity for jets with transverse momentum
above 30 GeV and $|\eta|\leq 4.9$. From the figure we would choose events with either one or two jets. 
However, and again due to the QCD background, we have chosen events  with one and one jet only.

\item In figure~\ref{fig5} (right) we present the jet multiplicity after the previous cuts, for jets with $p_T>15$ GeV and $|\eta|\leq 4.9$.
Taking into account the previous cuts, this is the jet multiplicity for jets with transverse momentum 
between 15 and 30 GeV. We veto all events with a jet multiplicity equal to two or above.

\begin{figure}[here]
  \begin{center}
    \includegraphics[scale=0.37]{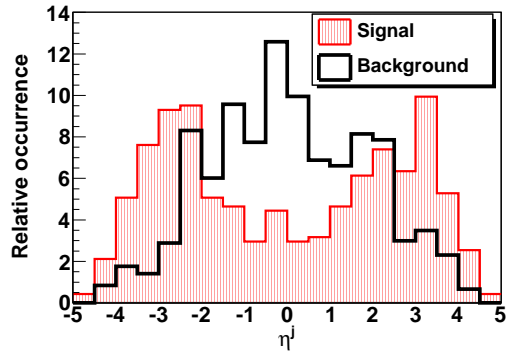}
    \caption{The jet pseudorapidity distribution. (Notice that after the previous cuts there is only one jet with $p_T>30 \: {\rm GeV}$ and $|\eta|\leq 4.9$.) }
    \label{fig6}
  \end{center}
\end{figure}

\item Finally, in  figure~\ref{fig6} we present the jet pseudorapidity distribution. The last cut is to accept events
where jets have a pseudorapidity $|\eta|\geq 2.5$.

\end{enumerate}

We have chosen a charged Higgs mass of 120 GeV to 
show the result of the analysis after all cuts have been applied (a complete cut flow is presented in Appendix A).
In table~\ref{tab:back} we list all backgrounds considered
in the analysis before and after cuts for a luminosity
of 1 fb$^{-1}$.  Clearly, after all cuts have been applied,  the main background is still
the single top background while the contributions $W + 2j$ and $W + 1j$ ($j={\rm jet}$) 
are the most important ones from the reducible background.
Given the very low efficiencies, the initial number of events generated
for each individual background
process was chosen in such a way that  the error in the total 
number of signal events is below 5\%. This is done by decreasing
the weight of each event until the required precision is attained.

In all backgrounds generated with AlpGen at LO, the renormalisation
and factorisation scales were chosen so that total cross section for
LO and NLO were similar~\cite{Campbell:2003hd}. After all cuts have been applied 
we have recalculated the LO and NLO total cross sections and found
that the $K$ factors were always of the order 1 or smaller.

\begin{table}
\begin{center}
\begin{tabular}{cccc}
\hline
& \quad Events (1 fb$^{-1}$) & Events (1 fb$^{-1}$) &\\
Process   & before cuts & after cuts &  Efficiency (\%)\\
\hline
Single top ($t$-channel)	 &			246600	 		                  & 18.5	& 0.0075 \\
Single top ($s$-channel)	 &			  10650		 	                  & 0 	    & 0 \\
Single top ($tW$)	             &			  66000		 	                  & 0.7 	    & 0.0010 \\
$t\bar t$ (semileptonic)	     &	         371133		 	              & 0   	& 0 \\
$t\bar t$ (leptonic)	         &		      88940	 		                  & 1.8 	& 0.0020 \\
$t\bar t$ (hadronic)	         &		    387169		 	                  & 0 	    & 0 \\
W+0j		                          &			4.3 $\times 10^7$    	 & 0 	    & 0 \\
W+1j		                         &			8.8 $\times 10^6$    	 & 0.6 	    & $7.1\times 10^{-6}$ \\
W+2j		                         &			2.8 $\times 10^6$    	 & 3.9 	& 0.00014 \\
W+3j		                         &			1.2 $\times 10^6$  	 	 & 0 	    & 0 \\
Wc+0j		                         &			7.4 $\times 10^5$    	 & 0 	    & 0 \\
Wc+1j		                         &			3.2 $\times 10^5$    	 & 3.2 	& 0.0010 \\
Wc+2j		                         &			2.0 $\times 10^5$    	 & 1.0 	    & 0.0011 \\
Wc+3j		                         &			8.9 $\times 10^4$  	 	 & 0 	    & 0 \\
Wbb+0j		                    &			6638                         	 & 0 	    & 0 \\
Wbb+1j		                    &			6582                         	 & 0.5 	    & 0.007 \\
Wbb+2j		                    &			3746    	                          & 0 	    & 0 \\
Wbb+3j		                    &			2896                     	 	 & 0 	    & 0 \\
\hline
\end{tabular}
\end{center}
\caption{Number of background events before and after cuts, alongside the same rates for the signal assuming a charged Higgs mass of 120 GeV.}
\label{tab:back}
\end{table}

In table~\ref{tab:signal} we show the number of events before and after all cuts
as a function of the charged Higgs mass. 
Recall that signal means all the events produced in $pp \to tj \to b H^{\pm} j $
with $H^{\pm}  \to \tau \nu \to l \slashed E$, where $l$ is a lepton
 and $\slashed E$ is the transverse missing energy. However, in order,
 to use these numbers in a variety of models (to be discussed later on)
 we have taken BR$ (t \to b H^{\pm}) = 100 \%$ and 
BR$ (H^- \to \tau^- \nu) = 100 \%$ and all other
BRs have the usual SM values. Also notice that 
the discontinuity in the number of
events after cuts for a mass of 130 GeV is due to the addition of the selection
cut  $M_T^{l \nu} > 85$ GeV for charged Higgs masses above 130 GeV.  
\begin{table}[here]
\begin{center}
\begin{tabular}{cccc}
\hline
$m_H^{\pm}$ (GeV)   &  Efficiency (\%)\\
\hline
90  	&	$0.016$\\
100 	&	$0.016$\\
110  & 	$0.018$\\
120  &  $0.019$\\
130  & $0.017$\\
140  &  $0.048$\\
150  &  $0.049$\\
160  &  $0.044$\\
%
%
%
\hline
\end{tabular}
\end{center}
\caption{Signal efficiency considering  BR$ (t \to b H^{\pm}) = 100 \%$ and 
BR$ (H^- \to \tau^- \nu) = 100 \%$ and all other
BRs have the usual SM values.} 
\label{tab:signal}
\end{table}

Putting all the numbers together we can find  $S/B$ and $S/\sqrt B$
as a function of the charged Higgs mass as presented in table~\ref{tab:signi}.
\begin{table}[h]
\begin{center}
\begin{tabular}{ccccc}
\hline
$m_H^{\pm}$ (GeV)   & Signal ($S$) & Background ($B$) & $S/B$ $(\%)$ & $S/\sqrt B$\\
\hline
90 & $38.6$ & $29.5$ & $130.92$ & $7.11$\\
100 & $40.5$ & $29.5$ & $137.19$ & $7.45$\\
110 & $45.6$ & $29.8$ & $153.00$ & $8.35$\\
120 & $47.7$ & $30.1$ & $158.26$ & $8.69$\\
130 & $42.3$ & $32.68$ & $129.53$ & $7.41$\\
140 & $117.1$ & $77.9$ & $150.25$ & $13.26$\\
150 & $120.0$ & $86.6$ & $138.64$ & $12.90$\\
160 & $109.7$ & $100.8$ & $108.81$ & $10.92$\\ 
\hline
\end{tabular}
\end{center}
\caption{Signal-to-Background ratio ($S/B$) 
and significance ($S/\sqrt B$) as a function of the charged Higgs mass.
The 
numbers presented for the signal we take BR$ (t \to b H^{\pm}) = 100 \%$ and 
BR$ (H^- \to \tau^- \nu) = 100 \%$ and all other
BRs have the usual SM values.} 
\label{tab:signi}
\end{table}

Just to give an idea how these numbers are modified for the different
models we have chosen two reference points: one for model II, with
$\tan \beta = 1$ and one for model type X for $\tan \beta = 3$. In the first
case we obtain $S/\sqrt B = 1.4$ and $S/B = 25 \%$ while in the second
the values are $S/\sqrt B = 0.3$ and $S/B = 5 \%$. How the values of the branching ratios
affect the results in the different models will be shown in the exclusion
plots presented in the next section.

\section{Results and discussion}

\begin{figure}[h!]
  \begin{center}
    \includegraphics[scale=0.37]{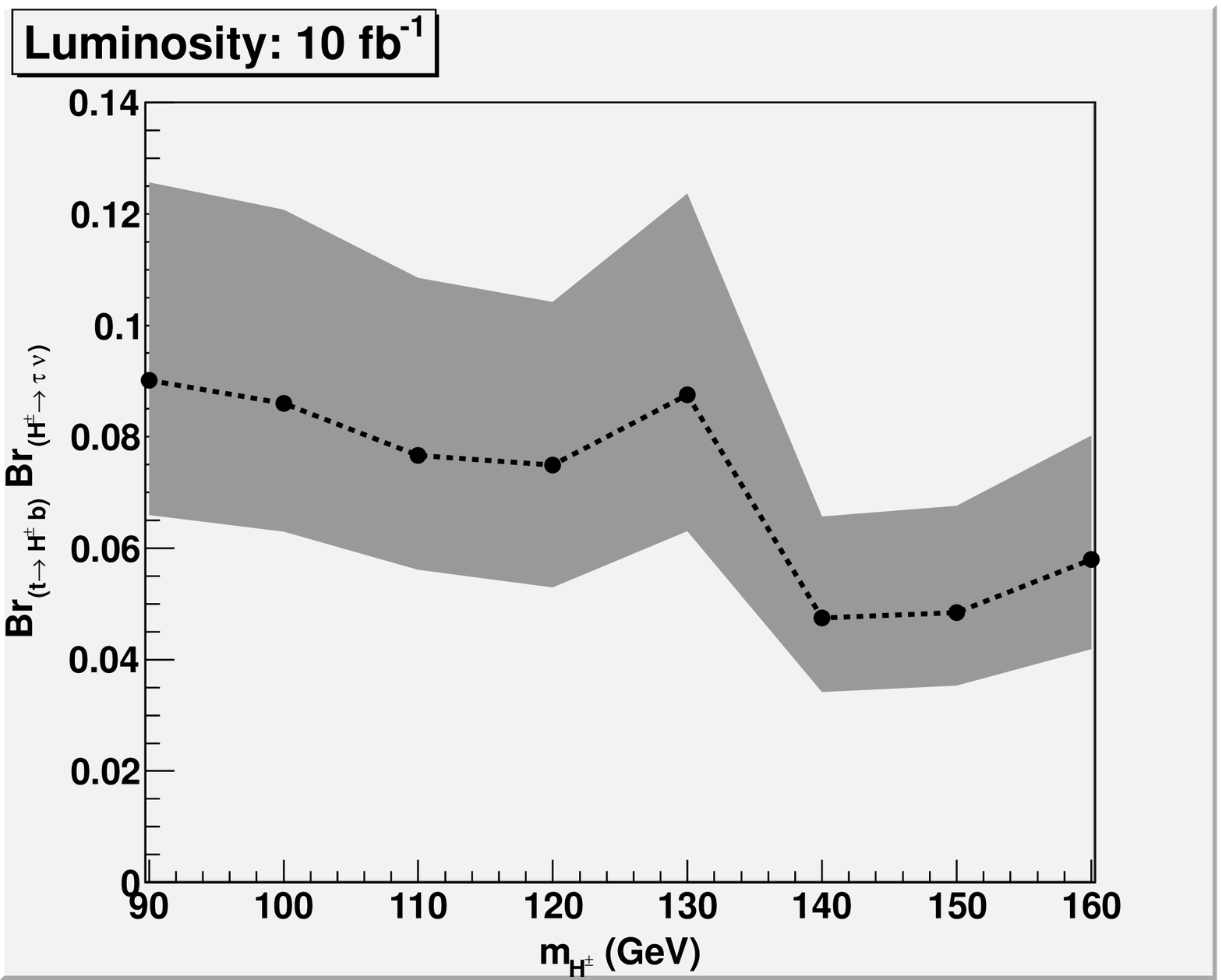}
    \includegraphics[scale=0.37]{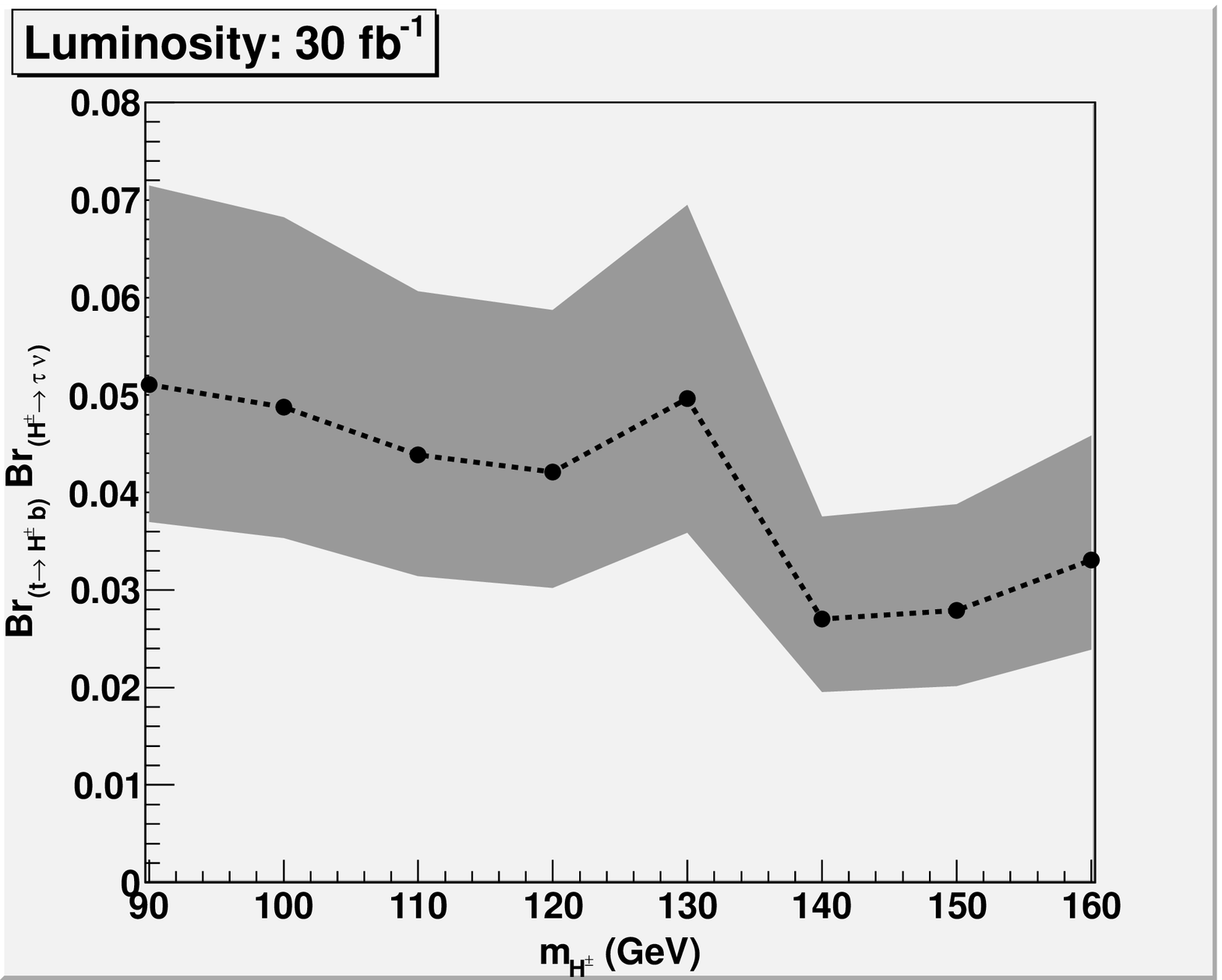}
    \caption{Left: exclusion limits for $ {\rm BR}{(t(\bar t)\rightarrow H^\pm b)}\: {\rm BR}{(H^\pm\rightarrow \tau^\pm \nu_{\tau})}$ 
    as a function of the charged Higgs mass at 95\% CL for 10 fb$^{-1}$ of integrated luminosity. 
    Right: exclusion limits for $ {\rm BR}{(t(\bar t)\rightarrow H^\pm b)}\: {\rm BR}{(H^\pm\rightarrow \tau^\pm \nu_{\tau})}$
     as a function of the charged Higgs mass at 95\% CL for 30 fb$^{-1}$ of integrated luminosity.}
    \label{fig7}
  \end{center}
\end{figure}

We start by presenting our results in a model independent manner, where only  the $\tau$ 
lepton decays according to the SM branching fractions are considered. Besides, we also
consider the top production cross section to be the SM one and because the results are presented for
a definite collected luminosity, the exclusion limits are presented for 
$ {\rm BR}{(t(\bar t)\rightarrow H^\pm b)}\: {\rm BR}{(H^\pm\rightarrow \tau^\pm \nu_{\tau})}$.
To obtain the 95\% CL limits we use
a code briefly described in~\cite{Veloso:2008zza} which is based on a ROOT library.
In the left panel of figure~\ref{fig7} we show the
 95\% CL limit for $ {\rm BR}{(t(\bar t)\rightarrow H^\pm b)}\: {\rm BR}{(H^\pm\rightarrow \tau^\pm \nu_{\tau})}$
as a function of the charged Higgs mass with an integrated luminosity of 10 {fb$^{-1}$} and for $\sqrt s = 14$ TeV.
In the same figure on the right, we show the results for an integrated luminosity of 30 {fb$^{-1}$} instead.

\begin{figure}[here]
  \begin{center}
    \includegraphics[scale=0.37]{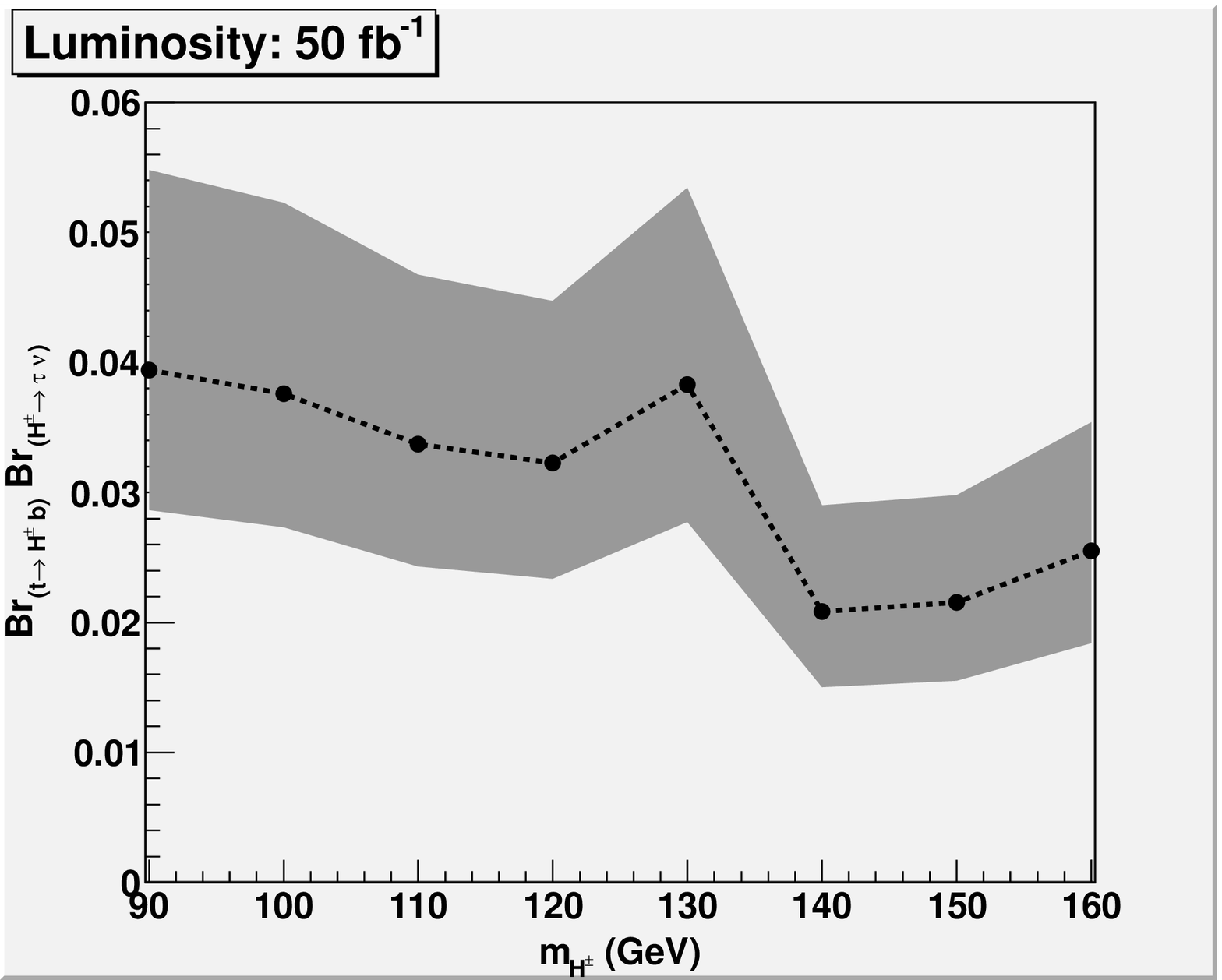}
    \includegraphics[scale=0.37]{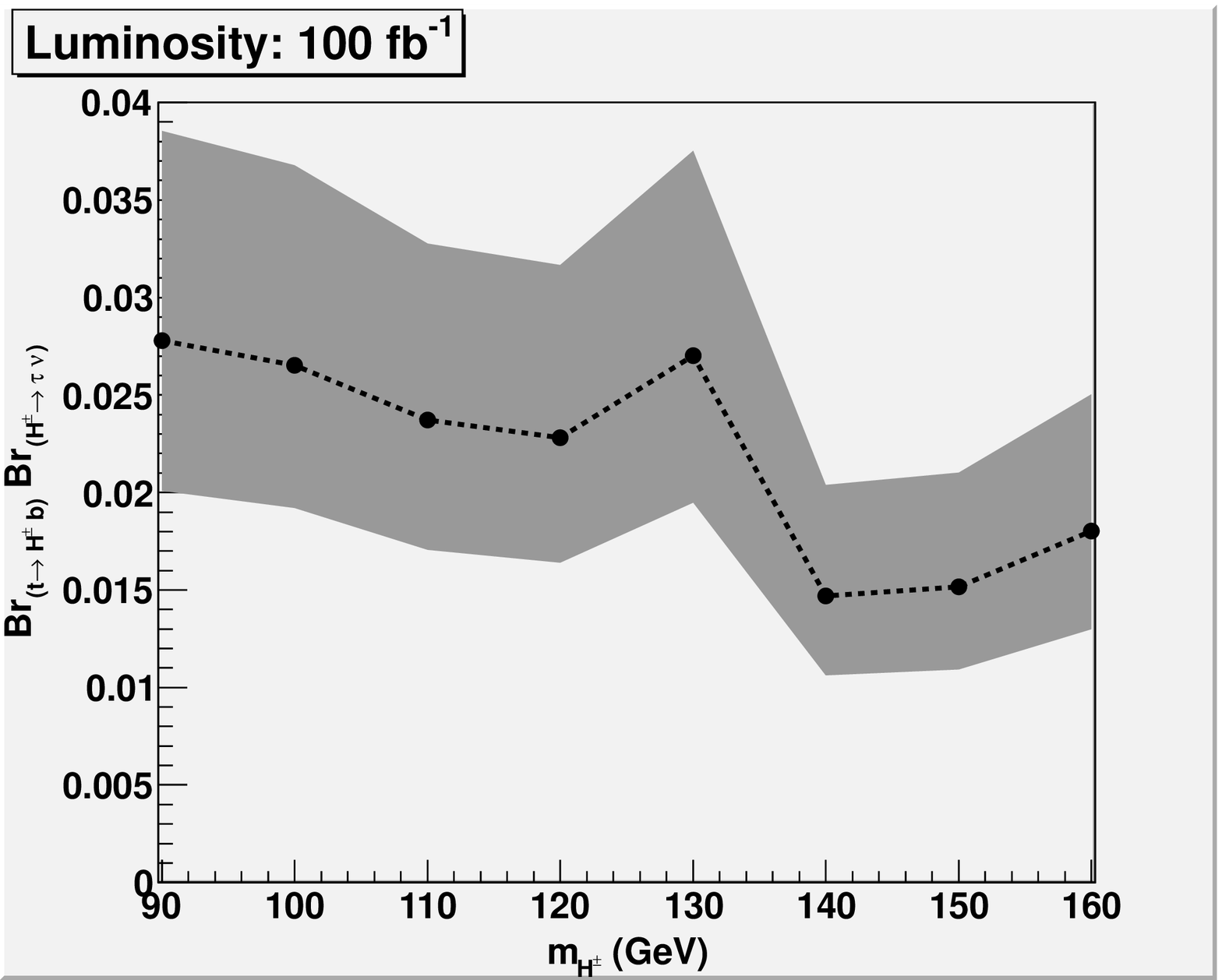}
    \caption{Left: exclusion limits for ${\rm BR}{(t(\bar t)\rightarrow H^\pm b)}\: {\rm BR}{(H^\pm\rightarrow \tau^\pm \nu_{\tau})}$ 
    as a function of the charged Higgs mass at 95\% CL and 50 fb$^{-1}$ of integrated luminosity. 
    Right: exclusion limits for ${\rm BR}{(t(\bar t)\rightarrow H^\pm b)}\: {\rm BR}{(H^\pm\rightarrow \tau^\pm \nu_{\tau})}$ 
    as a function of the charged Higgs mass at 95\% CL and 100 fb$^{-1}$ of integrated luminosity.}
    \label{fig8}
  \end{center}
\end{figure}

%
In figure~\ref{fig8} we present the same limits 
for  an integrated luminosity of 50 {fb$^{-1}$} (left) and  100 {fb$^{-1}$} (right). 

%

The results presented for $ {\rm BR}{(t(\bar t)\rightarrow H^\pm b)}\: {\rm BR}{(H^\pm\rightarrow \tau^\pm \nu_{\tau})}$ 
can be used to constrain any model where the the top decays to a charged Higgs boson which subsequently decays to $\tau^\pm \nu_{\tau}$.
The predicted exclusion bounds can be compared to similar plots  presented by the ATLAS~\cite{ATLASICHEP} 
and CMS~\cite{CMSICHEP} collaborations 
for the 7 TeV run and for the $t \bar t$ mode. The ATLAS (CMS) collaboration has an exclusion that ranges from 5 (4) \% to a charged 
Higgs mass of 90 GeV to 1 (2) \% for a mass of 160 GeV. We note once more that the search for a charged Higgs
in single top production does not and cannot compete with the $t \bar t$ search. Its purpose is to be combined
with the $t \bar t$ results to improve the sensitivity in charged Higgs boson searches. 


There are several models that we can explore now. Both the MSSM
and several versions of 2HDMs have a similar Yukawa Lagrangian in what concerns
the charged Higgs boson couplings to fermions. In fact, 2HDM types with a softly broken $Z_2$ symmetry, $\Phi_1 \rightarrow \Phi_1$,
$\Phi_2 \rightarrow - \Phi_2$, either
CP-conserving or CP-violating (either explicit or spontaneous) can be written in general terms as
\begin{eqnarray}
V(\Phi_1,\Phi_2) &=& m^2_1 \Phi^{\dagger}_1\Phi_1+m^2_2
\Phi^{\dagger}_2\Phi_2 + (m^2_{12} \Phi^{\dagger}_1\Phi_2+{\rm
h.c}) +\frac{1}{2} \lam_1 (\Phi^{\dagger}_1\Phi_1)^2 +\frac{1}{2}
\lam_2 (\Phi^{\dagger}_2\Phi_2)^2\nonumber \\ &+& \lam_3
(\Phi^{\dagger}_1\Phi_1)(\Phi^{\dagger}_2\Phi_2) + \lam_4
(\Phi^{\dagger}_1\Phi_2)(\Phi^{\dagger}_2\Phi_1) + \frac{1}{2}
\lam_5[(\Phi^{\dagger}_1\Phi_2)^2+{\rm h.c.}] ~, \label{higgspot}
\end{eqnarray}
where $\Phi_i$, $i=1,2$ are complex SU(2) doublets with four degrees of freedom each.
The parameters $m_{12}^2$,  $\lambda_5$ and the nature
of the VEVs will determine the CP nature
of the model (see~\cite{Branco:2011iw} for a review on the different 2HDMs). 
Note that hermiticity of the potential
forces the remaining parameters to be real. This, in turn,
will give rise to different neutral scalar sectors: if CP
is conserved we end up with two CP-even Higgs states, 
usually denoted by $h$ and $H$, and one CP-odd state, usually denoted by $A$;
otherwise we will just have three spinless states with undefined CP quantum numbers usually denoted
by $h_1$,  $h_2$ and $h_3$. However, as long as
the VEV does not break the electric charge, which was shown
to be possible in any 2HDM~\cite{vacstab1}, 
there are in any case two (identical) charged Higgs boson states, one charged conjugated to the other.

For definiteness, we will concentrate on two specific realisations, one CP-con\-serv\-ing
and the other explicitly CP-violating~\cite{Ginzburg:2002wt,ElKaffas:2006nt,Arhrib:2010ju,Barroso:2012wz},  and both are free from 
tree-level FCNCs. In the CP-violating version
$m_{12}^2$ and $\lambda_5$ are complex and $Im (\lambda_5) = 2 ~Im (m_{12}^2)$.
In both models the VEVs are real. By defining 
$\tan\beta=v_2/v_1$, it is then possible to choose the angle $\beta$
 as the rotation angle from the group eigenstates to the mass eigenstates in the 
 charged Higgs sector.
 By then extending the $Z_2$ symmetry to the Yukawa sector 
we end up with four independent 2HDMs,
i.e., the aforementioned Type I, Type II, Type Y and Type X,
 whose $H^\pm$ couplings to fermions are presented in table~\ref{tab:Yuk}.  
Herein, $P_L$ and $P_R$ are the left- and right-helicity projection operators, respectively. 
\begin{table}
\begin{center}
\begin{tabular}{ccc}
\hline
Model   & $g_{\bar u d H^+}$ & $g_{ l \bar \nu  H^+}$    \\
\hline
I  & $\frac{ig}{\sqrt 2 \, M_W} V_{ud} \, [- m_d/\tan \beta P_R + m_u/\tan \beta P_L]$	& $\frac{ig}{\sqrt 2 \, M_W} \,  [- m_l/\tan \beta P_R]$	 \\
II & $\frac{ig}{\sqrt 2 \, M_W} V_{ud} \, [m_d \, \tan \beta P_R + m_u/\tan \beta P_L]$	& $\frac{ig}{\sqrt 2 \, M_W} \,  [m_l \, \tan \beta P_R]$	\\
Y & $\frac{ig}{\sqrt 2 \, M_W} V_{ud} \, [m_d \, \tan \beta P_R + m_u/\tan \beta P_L]$	& $\frac{ig}{\sqrt 2 \, M_W}  \,  [- m_l/\tan \beta P_R]$	 \\
X & $\frac{ig}{\sqrt 2 \, M_W} V_{ud} \, [- m_d/\tan \beta P_R + m_u/\tan \beta P_L]$	& $\frac{ig}{\sqrt 2 \, M_W} \,  [m_l \, \tan \beta P_R]$	\\
\hline
\end{tabular}
\end{center}
\caption{Charged Higgs Yukawa couplings to up-, down-type quarks and leptons.} 
\label{tab:Yuk}
\end{table}
Further, notice that the vertices  $g_{u \bar d H^-}$ and $g_{\bar  l \nu  H^-}$   are obtained from the corresponding  
ones $g_{\bar u d H^+}$ and  $g_{ l \bar \nu  H^+}$ by interchanging $P_L \leftrightarrow P_R$ and 
by replacing $V_{ud} \rightarrow V_{ud} ^*$.  

With this parametrisation the results presented here depend only on $\tan \beta$ and 
$m_{H^\pm}$. This is true both for BR${(t(\bar t)\rightarrow H^\pm b)}$
and BR${(H^\pm\rightarrow \tau^\pm \nu_{\tau})}$. The experimental
bounds on the CP-violating model, which for the charged sector
are also valid for the CP-conserving one, were recently reviewed in~\cite{Basso:2012st}.
Taking into account these bounds we have finally looked for the 95\% CL exclusion
limits in each model.

\begin{figure}[here]
  \begin{center}
    \includegraphics[scale=0.37]{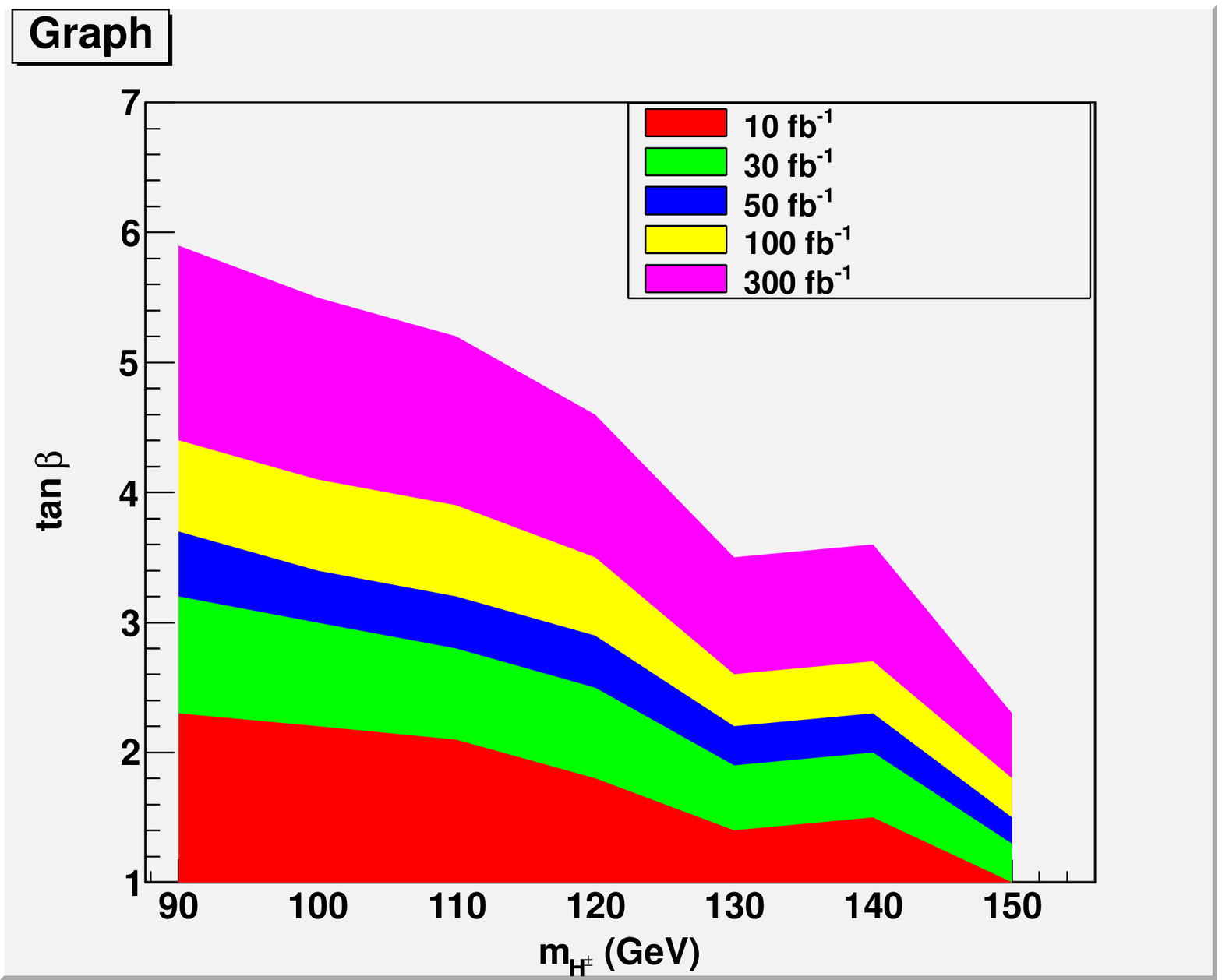}
    \includegraphics[scale=0.37]{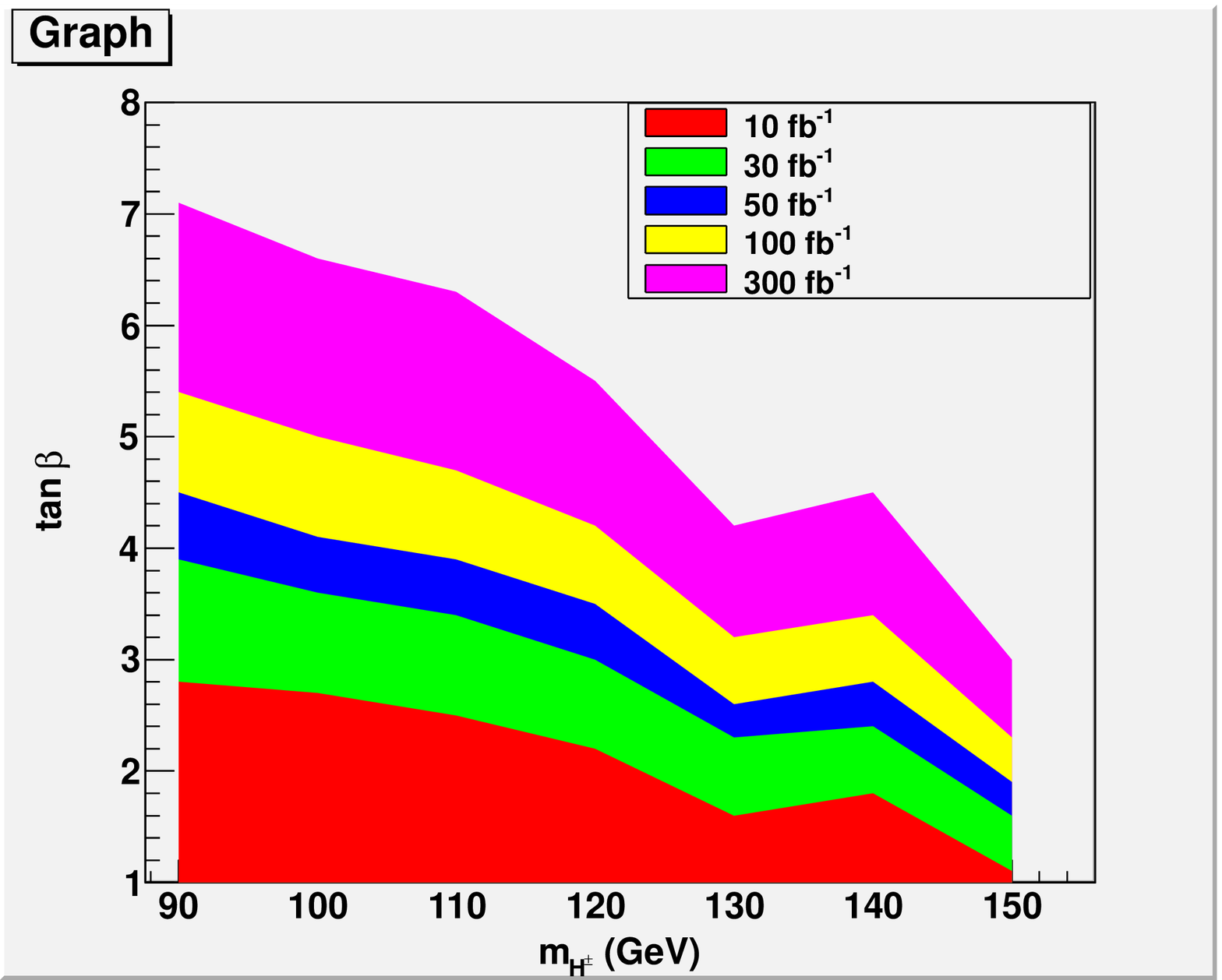}
    \caption{The 2HDM Type I (left) and Type X (right) exclusion limits over the ($\tan\beta$, $m_{H^\pm}$) plane at the 95\% CL
    assuming the LHC at 14 TeV and for several luminosity sets.}
    \label{fig10}
  \end{center}
\end{figure}

In figure~\ref{fig10} we present the  95\% CL exclusion
limits for a 2HDM Type I (left) and Type X (right). These results should now be compared
with the predictions for $\sqrt{s} = 14$ TeV  made by the experimental collaborations
at CERN~\cite{Aad:2009wy, CMScharged}. Using the ATLAS analysis presented in~\cite{Aad:2009wy} we can draw 
exclusion plots for the different models as presented in~\cite{Aoki:2011wd}. Using Type
I as an example, ATLAS sets a limit $\tan \beta < 9$ for a collected luminosity of 30 fb$^{-1}$
and for a charged Higgs mass of $m_{H^\pm} = 90$ GeV. Figure~\ref{fig10} shows
that for the same luminosity we obtain a limit slightly above 3 for $\tan \beta$. For other
masses the result is slightly better but the general trend is a factor between 2 and 3 in the ratio
of the limits obtained in the two analysis. It should be noted that the ATLAS analysis considers both
the leptonic and hadronic decays of the $\tau$ leptons reaching, also for that reason,
a much higher sensitivity.

\begin{figure}[here]
  \begin{center}
    \includegraphics[scale=0.37]{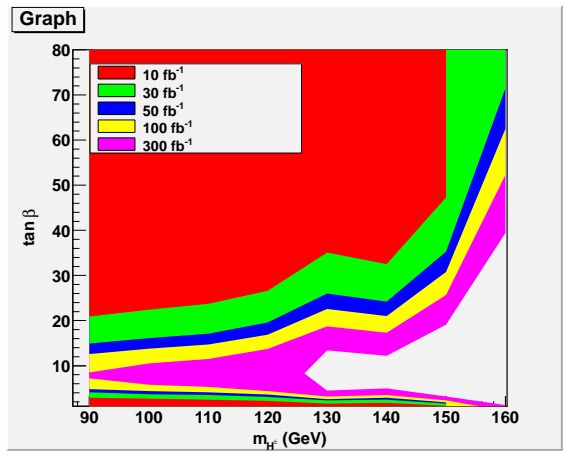}
    \caption{The MSSM exclusion limits over the ($\tan\beta$, $m_{H^\pm}$) plane at the 95\% CL.
    assuming the LHC at 14 TeV and for several luminosity sets.}
    \label{fig11}
  \end{center}
\end{figure}

In figure~\ref{fig11} we present the  95\% CL exclusion
limits for a 2HDM Type II, which for phenomenological purposes we may assume it to be the MSSM, so as have the light charged Higgs boson considered here compliant
with the $b\to s\gamma$ constraints. Again these results should be compared with the 14 TeV predictions
for the LHC in the $t \bar t$ production channel. For comparison purposes, let us consider
figure~\ref{fig11} and the exclusion region for 30 fb$^{-1}$. The allowed region for 
$m_{H^\pm} = 90$ GeV is for $4.1 < \tan \beta < 14.9$ while our best 
result is for $m_{H^\pm} = 140$ GeV where the allowed range of $\tan \beta$ at 95\% CL is $2.3 < \tan \beta < 24.2$.
The ATLAS prediction~\cite{Aad:2009wy} for the same energy and for the same collected
luminosity and considering only the the leptonic decays of the $\tau$ lepton  
is $7.1 < \tan \beta < 9.3$  for $m_{H^\pm} = 90$ GeV and is $3.2 < \tan \beta < 21.9$  for $m_{H^\pm} = 130$ GeV
at 95\% CL. Therefore, it is clear that the results obtained in the single top analysis should
be combined with the ones from $t \bar t$ production to improve the sensitivity. We note however that
this comparison is made with the leptonic mode in the $t \bar t$ and when all the $\tau$
decay modes are combined the result obviously improves as presented in~\cite{Aad:2009wy}.
  
Finally, in comparing the results in figures~\ref{fig10} and \ref{fig11}, it should be noticed that
that there is sensitivity up 
to $\tan\beta= 70$ in the 2HDM Type II (herein the MSSM) whereas this is only up to
7 in the case of a 2HDM Type I or Type IV. This is because 
in the Type I and Type IV case the cross section times BR rates always decrease with an increase
of $\tan\beta$, while in Type II the cross section times BR rates decrease with
an increase of $\tan\beta$ from 1 to some point, 6 or 7, then they start to
increase again.

\section{Summary, conclusions and outlook}

Now that a neutral Higgs boson signal may be about to be found at the LHC, it becomes of paramount importance to assess
the dynamics of the underlying Higgs mechanism of EWSB. In fact, early data (and possibly late ones to appear soon) seem
to hint that the excess is above the SM expectations, at least in the (most prominent, for the extracted mass of $\approx 125$ GeV) 
$\gamma\gamma$ decay mode. Therefore, a non-minimal Higgs sector may be required. If the latter involves at least another Higgs doublet,
then a (singly) charged Higgs boson is bound to appear in the model spectrum. Of particular phenomenological relevance are 2HDMs,
where one and only one complex Higgs doublet is added to the SM one. Herein, there is only one charged Higgs boson state and its
mass can be predicted (through unitarity) to be well within the kinematic reach of the LHC. In fact, it is the low mass regime,
i.e., $m_{H^\pm}<m_t$, which is the spectacular one, as such a light charged Higgs boson can be generated in top-(anti)quark decays,
since the latter are copiously produced at the LHC through efficient QCD interactions.

While past literature exploring the scope of the LHC in the search for light charged Higgs bosons in 2HDMs has concentrated exclusively on the case of $H^\pm$ production from $t$-(anti)quarks
produced in pairs, we have tackled here the case of single-top production, whose cross section is in fact comparable (albeit smaller)
to that of double-top production, at all LHC energies. We have found that, when searched for in leptonic decays of primary $\tau\nu$ pairs, the LHC is very sensitive
to such a light charged Higgs boson, in the most common (and still viable) realisations of 2HDMs, including the one whose Higgs sector is naturally embedded in Supersymmetry, the so-called
MSSM. We have in fact quantified this in terms of exclusion regions, particularly in the MSSM, and proved that single-top production offers 
an alternative
means (albeit of more limited scope) to double-top production to test the existence of otherwise of light $H^\pm$ states. This is most 
efficiently done at 14 TeV, 
assuming standard luminosity samples. In this respect, the main highlights 
are that our results are a factor between 2 and 3 below the ones obtained 
by ATLAS and CMS for the $t \bar t$ production channel. Therefore they
can be combined to provide improved exclusion regions for the parameter
space of the models. Although not presented by ATLAS and CMS we can state
that regarding discovery 
even for 300 fb$^{-1}$ and for a charged Higgs
mass of 100 GeV we can barely reach $\tan \beta =3$ for Types I and X (and lower limit for Type II) while for higher masses
the discovery limit is only possible for a $\tan \beta \,  {\cal O}(1)$.
Our conclusions are supported by a thorough phenomenological investigation of signal and both reducible and irreducible backgrounds, in presence of PS, hadronisation and detector
effects.

Having proved the accessibility of charged Higgs production in single top mode at the LHC, possible outlooks include now the following.
\begin{enumerate}
\item The investigation of the Cabibbo-Kobayashi-Maskawa structure of the $H^\pm tb$ vertex, which in single top mode appears at production level,
unlike in the double top  channel where such a vertex only enters the top decay, thereby affording more sensitivity to experimental measurements.
\item Because of such a sensitivity at production level to this coupling, one can also attempt extracting the coefficients of the two chiral structures
$(1\pm\gamma^5)/2$ entering the $H^\pm tb$ vertex, by studying the angular behaviour of the final state particles entering single top production
(recall in fact that the inclusive cross section is only proportional to the sum of the squares of such coefficients).
\end{enumerate}
These aspects will constitute the subject of a separate publication.      

\section*{Acknowledgments}
We thank Michelangelo Mangano for his help
with AlpGen and for discussions. We thank Emanuele Re
for his help with POWHEG (especially for
modifying the program to allow the generation
without the top decaying) and for discussions. We thank Filipe Veloso for
providing us with the code to extract the exclusion limits and discussions.
We thank Miguel Won for help with stdhep. We thank Nuno Castro for discussions.

SM is financed in part through the NExT Institute. 
The work of RG and RS is supported in part by the Portuguese
\textit{Funda\c{c}\~{a}o para a Ci\^{e}ncia e a Tecnologia} (FCT)
under contracts PTDC/FIS/117951/2010 and PEst-OE/FIS/UI0618/2011.
RG is also supported by a FCT Grant SFRH/BPD/47348/2008.
RS is also partially supported by an FP7 Reintegration Grant, number PERG08-GA-2010-277025.

\appendix
\section{Cut flow for a 120 GeV charged Higgs boson}

In table~\ref{tab:flow1} we present the cut flow for the signal ($m_{H^{\pm} }$ = 120 GeV)
and for the background starting from the trigger until cut number 5 as described in section
3. In table~\ref{tab:flow2}  we show the remaining cuts numbered from 6 to 10. All
numbers are efficiencies in percentage. 

\begin{table}[here]
\begin{center}
\begin{tabular}{c|cccccc}
\hline
& 	\multicolumn{6}{c}{Efficiency after cut (\%)}	\\
Process	&	Trigger	&	Cut 1	&	Cut 2	&	Cut 3	&	Cut 4	&	Cut 5\\
\hline
Signal &	45.57	&	 9.36	&	 8.32	&	 6.86	&	 3.46	&	1.00\\
& 	\multicolumn{6}{c}{}	\\
tj ($t$)				&	54.73	&	13.57	&	11.80	&	 8.61	&	 3.60	&	1.00\\
tj ($s$)				&	59.01	&	14.87	&	12.31	&	 8.91	&	 3.58	&	0.77\\
tj ($tW$)				&	80.62	&	25.89	&	21.22	&	11.74	&	 5.93	&	1.34\\
$t\bar t$ (sem)	&	71.15	&	38.34	&	30.47	&	17.11	&	10.63	&	2.57\\
$t\bar t$ (lep)	&	84.07	&	43.45	&	26.51	&	14.87	&	11.04	&	2.59\\
$t\bar t$ (had)	&	28.22	&	 3.39	&	 2.53	&	 2.07	&	 0.34	&	0.05\\
W+0j						&	26.14	&	23.50	&	23.50	&	23.36	&	 1.05	&	0.01\\
W+1j						&	34.00	&	27.73	&	27.65	&	21.83	&	 4.85	&	0.09\\
W+2j						&	38.89	&	28.27	&	28.08	&	19.50	&	 5.83	&	0.16\\
W+3j						&	48.10	&	29.08	&	28.71	&	17.29	&	 7.16	&	0.27\\
Wc+0j						&	41.20	&	35.96	&	35.18	&	25.48	&	 4.29	&	0.35\\
Wc+1j						&	45.01	&	36.60	&	35.55	&	22.69	&	 5.40	&	0.43\\
Wc+2j						&	50.37	&	37.46	&	36.07	&	20.34	&	 6.72	&	0.54\\
Wc+3j						&	57.04	&	37.95	&	36.16	&	18.16	&	 7.57	&	0.53\\
Wbb+0j					&	40.70	&	31.01	&	27.35	&	19.43	&	 5.46	&	1.69\\
Wbb+1j					&	44.71	&	28.74	&	25.33	&	15.48	&	 5.83	&	1.76\\
Wbb+2j					&	51.02	&	29.34	&	25.78	&	15.22	&	 7.56	&	2.45\\
Wbb+3j					&	23.15	&	10.99	&	 9.17	&	 4.34	&	 3.08	&	5.98$\times10^{-1}$\\
\hline
\end{tabular}
\end{center}
\caption{Efficiency flow (trigger  to cut 5) for a charged Higgs mass of 120 GeV.}
\label{tab:flow1}
\end{table}

\begin{table}[here]
\begin{center}
\begin{tabular}{c|cccccc}
\hline
& 	\multicolumn{6}{c}{Efficiency after cut (\%)}	\\
Process	&		&	Cut 6	&	Cut 7	&	Cut 8	&	Cut 9	&	Cut 10\\
\hline
Signal&		&	4.33$\times10^{-1}$	&	1.13$\times10^{-1}$	&	4.36$\times10^{-2}$	&	3.94$\times10^{-2}$	&	1.93$\times10^{-2}$\\
& 	\multicolumn{6}{c}{}	\\
tj ($t$)				&		&	2.62$\times10^{-1}$	&	6.10$\times10^{-2}$	&	1.62$\times10^{-2}$	&	1.47$\times10^{-2}$	&	7.50$\times10^{-3}$\\
tj ($s$)				&		&	1.43$\times10^{-1}$	&	3.67$\times10^{-2}$	&	8.33$\times10^{-2}$	&	6.67$\times10^{-3}$	&	$\approx 0$      		\\
tj ($tW$)				&		&	4.58$\times10^{-1}$	&	1.31$\times10^{-1}$	&	1.10$\times10^{-3}$	&	6.00$\times10^{-3}$	&	1.00$\times10^{-3}$\\
$t\bar t$ (sem)	&		&	1.09               	&	2.93$\times10^{-1}$	&	9.12$\times10^{-2}$	&	4.50$\times10^{-3}$	&	$\approx 0$      		\\
$t\bar t$ (lep)	&		&	7.31$\times10^{-1}$	&	1.75$\times10^{-1}$	&	1.95$\times10^{-3}$	&	1.45$\times10^{-2}$	&	2.00$\times10^{-3}$\\
$t\bar t$ (had)	&		&	4.29$\times10^{-2}$	&	3.50$\times10^{-3}$	&	$\approx 0$      		&	$\approx 0$      		&	$\approx 0$      		\\
W+0j						&		&	1.48$\times10^{-4}$	&	$\approx 0$					&	$\approx 0$      		&	$\approx 0$      		&	$\approx 0$      		\\
W+1j						&		&	3.20$\times10^{-3}$	&	7.22$\times10^{-4}$	&	3.36$\times10^{-4}$	&	3.07$\times10^{-4}$	&	7.15$\times10^{-6}$\\
W+2j						&		&	2.41$\times10^{-2}$	&	5.88$\times10^{-3}$	&	2.33$\times10^{-3}$	&	2.19$\times10^{-3}$	&	1.39$\times10^{-4}$\\
W+3j						&		&	9.51$\times10^{-2}$	&	2.15$\times10^{-2}$	&	7.58$\times10^{-4}$	&	5.89$\times10^{-4}$	&	$\approx 0$      		\\
Wc+0j						&		&	3.93$\times10^{-3}$	&	5.74$\times10^{-4}$	&	8.19$\times10^{-5}$	&	8.19$\times10^{-5}$	&	$\approx 0$      		\\
Wc+1j						&		&	4.63$\times10^{-2}$	&	7.33$\times10^{-3}$	&	3.54$\times10^{-3}$	&	3.54$\times10^{-3}$	&	1.01$\times10^{-3}$\\
Wc+2j						&		&	1.18$\times10^{-1}$	&	2.45$\times10^{-2}$	&	5.32$\times10^{-3}$	&	2.13$\times10^{-3}$	&	1.06$\times10^{-3}$\\
Wc+3j						&		&	1.92$\times10^{-1}$	&	5.49$\times10^{-2}$	&	2.74$\times10^{-3}$	&	$\approx 0$      		&	$\approx 0$      		\\
Wbb+0j					&		&	1.31$\times10^{-1}$	&	2.95$\times10^{-2}$	&	2.11$\times10^{-2}$	&	2.11$\times10^{-2}$	&	$\approx 0$      		\\
Wbb+1j					&		&	3.35$\times10^{-1}$	&	1.05$\times10^{-1}$	&	4.19$\times10^{-2}$	&	3.49$\times10^{-2}$	&	6.98$\times10^{-3}$\\
Wbb+2j					&		&	7.80$\times10^{-1}$	&	1.89$\times10^{-1}$	&	$\approx 0$      		&	$\approx 0$      		&	$\approx 0$      		\\
Wbb+3j					&		&	2.99$\times10^{-1}$	&	6.90$\times10^{-2}$	&	$\approx 0$      		&	$\approx 0$      		&	$\approx 0$\\
\hline
\end{tabular}
\end{center}
\caption{Efficiency flow (cut 6  to cut 10) for a charged Higgs mass of 120 GeV.}
\label{tab:flow2}
\end{table}

\end{document}